\pdfoutput=1
\documentclass[sigconf]{acmart}
\AtBeginDocument{%
  \providecommand\BibTeX{{%
    \normalfont B\kern-0.5em{\scshape i\kern-0.25em b}\kern-0.8em\TeX}}}

\acmPrice{15.00}
\acmISBN{978-1-4503-XXXX-X/18/06}

\copyrightyear{2024}
\acmYear{2024}
\setcopyright{rightsretained}
\acmConference[KDD '24]{Proceedings of the 30th ACM SIGKDD Conference on
Knowledge Discovery and Data Mining}{August 25--29, 2024}{Barcelona, Spain}
\acmBooktitle{Proceedings of the 30th ACM SIGKDD Conference on Knowledge
Discovery and Data Mining (KDD '24), August 25--29, 2024, Barcelona,
Spain}\acmDOI{10.1145/3637528.3671924}
\acmISBN{979-8-4007-0490-1/24/08}

\makeatletter
\gdef\@copyrightpermission{
  \begin{minipage}{0.3\columnwidth}
   \href{https://creativecommons.org/licenses/by/4.0/}{\includegraphics[width=0.90\textwidth]{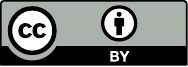}}
  \end{minipage}\hfill
  \begin{minipage}{0.7\columnwidth}
   \href{https://creativecommons.org/licenses/by/4.0/}{This work is licensed under a Creative Commons Attribution International 4.0 License.}
  \end{minipage}
  \vspace{5pt}
}
\makeatother

\usepackage[normalem]{ulem}
\useunder{\uline}{\ul}{}

\usepackage{amsmath}

\usepackage{balance}
\usepackage{pifont}
\usepackage{amsfonts, amsthm}
\usepackage[ruled,vlined,linesnumbered]{algorithm2e}
\usepackage{color, colortbl}
\usepackage{enumitem,multirow,graphicx,subcaption,multicol,lipsum,float,adjustbox}
\newlength{\textfloatsepsave}  
\usepackage{cleveref}
\usepackage{algorithmicx}
\usepackage[page]{appendix} 
\crefformat{section}{\S#2#1#3}
\crefformat{subsection}{\S#2#1#3}
\crefformat{subsubsection}{\S#2#1#3}

\begin{document}


\title{Continual Collaborative Distillation for Recommender System}

\begin{abstract}
Knowledge distillation (KD) has emerged as a promising technique for addressing the computational challenges associated with deploying large-scale recommender systems.
KD transfers the knowledge of a massive teacher system to a compact student model, to reduce the huge computational burdens for inference while retaining high accuracy.
The existing KD studies primarily focus on one-time distillation in static environments, leaving a substantial gap in their applicability to real-world scenarios dealing with continuously incoming users, items, and their interactions.
In this work, we delve into a systematic approach to operating the teacher-student KD in a non-stationary data stream.
Our goal is to enable efficient deployment through a compact student, which preserves the high performance of the massive teacher, while effectively adapting to continuously incoming data.
We propose \textbf{\underline{C}}ontinual \textbf{\underline{C}}ollaborative \textbf{\underline{D}}istillation (\textbf{CCD}) framework, where both the teacher and the student continually and collaboratively evolve along the data stream.
\proposed facilitates the student in effectively adapting to new data, while also enabling the teacher to fully leverage accumulated knowledge.
We validate the effectiveness of \proposed through extensive quantitative, ablative, and exploratory experiments on two real-world datasets.
We expect this research direction to contribute to narrowing the gap between existing KD studies and practical applications, thereby enhancing the applicability of KD in real-world systems.

\end{abstract}


\newcommand{\proposed}{CCD\xspace}
\newcommand{\tRS}{$RS^T$\xspace}
\newcommand{\sRS}{$RS^S$\xspace}

\newcommand{\smallsection}[1]{{\vspace{0.03in} \noindent \bf {#1}}}

\author{Gyuseok Lee}
\affiliation{
    \institution{Pohang University of \\ Science and Technology}
    \city{Pohang}
    \state{Gyeongbuk}
    \country{Republic of Korea}
}
\authornote{Both authors contributed equally to this research.}
\email{gyuseok.lee@postech.ac.kr}

\author{SeongKu Kang}
\affiliation{
    \city{}
    \institution{University of Illinois at \\ Urbana-Champaign}
    \city{Champaign}
    \state{IL}
    \country{USA}
}
\authornotemark[1]
\email{seongku@illinois.edu}

\author{Wonbin Kweon}
\affiliation{
    \institution{Pohang University of \\ Science and Technology}
    \city{Pohang}
    \state{Gyeongbuk}
    \country{Republic of Korea}
}
\email{kwb4453@postech.ac.kr}

\author{Hwanjo Yu}
\affiliation{
    \institution{Pohang University of \\ Science and Technology}
    \city{Pohang}
    \state{Gyeongbuk}
    \country{Republic of Korea}
}
\authornote{Corresponding author.}
\email{hwanjoyu@postech.ac.kr}

\maketitle

\section{Introduction}
Recommender systems (RS) have been used in diverse industrial platforms to enhance user experience, foster loyalty, and contribute to business success \cite{SSCDR, concf, lee2023mvfs}.
In recent years, increasingly large and sophisticated recommendation models, from graph neural networks \cite{ying2018graph, wang2020m2grl} to large language models \cite{geng2022recommendation, bao2023tallrec, Zhu_2024}, have been developed to identify users' intricate preferences.
Furthermore, these large-scale models are often employed concurrently through model ensembles to further improve recommendation performance \cite{hetcomp, zhu2020ensembled}. 
However, this improved performance comes at the cost of increased computations, memory resources, and inference latency, which poses significant challenges for deployment in real-time services and resource-constrained environments \cite{hetcomp, xia2022device, chen2022learning}.

\begin{figure}[t]
    \centering    
    \hspace{-0.2cm}
    \includegraphics[width=1.0\linewidth]{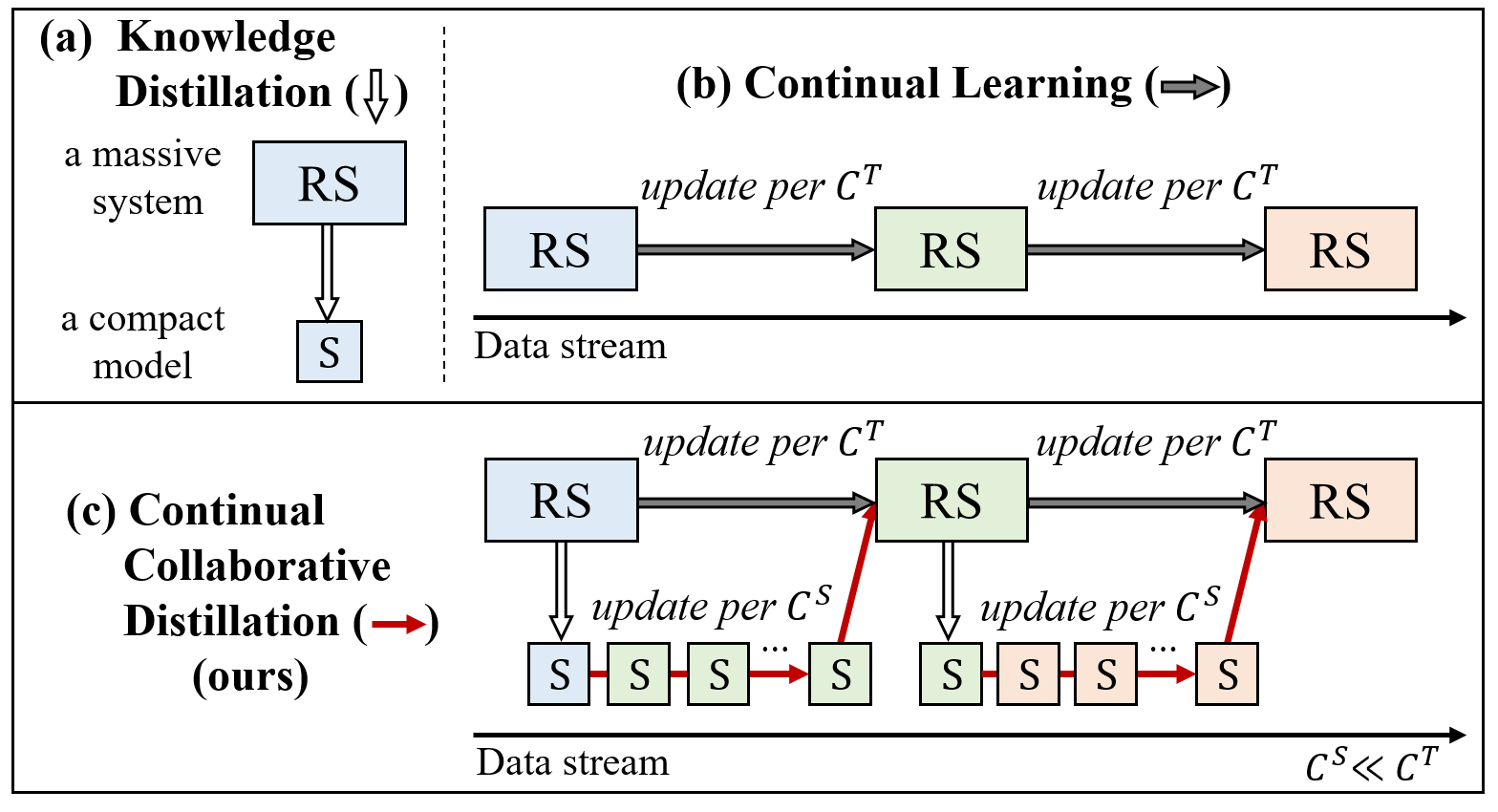}
    \caption{A conceptual comparison of (a) knowledge distillation, (b) continual learning, and (c) the proposed continual collaborative distillation. $C^T$ and $C^S$ denote the update cycle for the massive system (e.g., a weekly update) and the compact model (e.g., a daily update), respectively.}
    \label{fig:intro}
    \vspace{0.2cm}
\end{figure}

To overcome this issue, recent studies have employed knowledge distillation (KD) \cite{KD, FitNet} (Figure \ref{fig:intro}a).
KD serves as a model compression technique that transfers knowledge from a massive system (teacher) into a compact model (student), aiming to generate a model that achieves both high effectiveness and efficiency.
It first constructs a massive teacher system, which often comprises multiple large-scale models, to attain high performance \cite{hetcomp, zhu2020ensembled}.
Then, during the distillation process, the student is trained to replicate the high-quality recommendations from the teacher system, which are derived from its massive capacity.
After the distillation, the student becomes capable of achieving comparable performance to the teacher \cite{hetcomp, RD, xia2022device}.
Furthermore, the student has significantly reduced inference latency, making it suitable for deployment.

However, existing KD studies have focused on one-time distillation in static environments, overlooking real-world scenarios handling a non-stationary data stream where new users, items, and interactions are continuously incoming.
A naive approach to applying KD in the data stream is to repeatedly update the teacher and generate a new student via distillation each time new data arrives.
However, this approach raises two critical problems:
First, the massive teacher system cannot be updated frequently, as training large-scale models requires significant time and resources \cite{du2021alternate}. 
Consequently, the deployed student remains static until the next teacher update cycle, failing to provide recommendations that reflect up-to-date interactions as well as new users and items.
Second, conducting the distillation independently each time fails to fully leverage the previously generated knowledge along the data stream. 
This results in inefficient training and suboptimal performance.

A well-established approach for training a model with a non-stationary data stream is continual learning (CL) \cite{lin2022towards, kirkpatrick2017overcoming} (Figure \ref{fig:intro}b).
When trained with the data stream, an ideal model should seamlessly adapt to newly incoming data without forgetting previous knowledge from historical data.
CL aims to strike a balance between these two aspects, termed \textit{plasticity} (i.e., adaptation to new data) and \textit{stability} (i.e., preserving previous knowledge) \cite{lin2022towards}.
Although CL has been studied for RS \cite{LWCKD, PIW, zhu2023reloop2}, operating the teacher-student KD in a data stream still remains unexplored.
Most existing CL methods focus on the continual updates of \textit{a single model} with \textit{sufficient capacity} to learn effectively on its own.
However, KD necessitates handling both the teacher and the student which have distinct natures and update cycles.
Also, the student has a highly limited learning capability due to its small size, which makes it more challenging to update effectively.

We propose \textbf{\underline{C}}ontinual \textbf{\underline{C}}ollaborative \textbf{\underline{D}}istillation (\textbf{CCD}), a new approach to systematically operate the teacher-student KD in a non-stationary data stream (Figure \ref{fig:intro}c).
Our goal is to enable efficient deployment through a compact student which preserves the high performance of the massive teacher, while effectively adapting to continuously incoming data.
In \proposed, the student model, which can be updated frequently due to its small size, learns to adapt to new interactions with a short update cycle (e.g., a daily update).
Meanwhile, the teacher system uncovers richer knowledge through its larger capacity with a longer update cycle (e.g., a weekly update).
As shown in Figure \ref{fig:intro}c, within each teacher update cycle, we proceed through three consecutive stages:
(1) Compact student generation via KD,
(2) Continual student updates to provide recommendations reflecting up-to-date interactions,
(3) Teacher system update by leveraging knowledge accumulated along the data stream.

We introduce new effective update strategies for both the teacher and the student, considering their distinct natures.
To complement the student's limited learning capability, we propose two techniques: \textit{entity embedding initialization} which facilitates the learning of new users and items based on the recent prominent trends, and \textit{proxy-guided replay learning} which identifies forgotten knowledge and aids in its preservation by replaying past predictions.
Furthermore, we propose an effective teacher update strategy that fully leverages previously obtained model knowledge, including both from the teacher-side and the student-side.
By repeating the updating along the data stream, \proposed allows for collaboratively enhancing both the teacher and the student;
the student is enhanced by effectively adapting to new entities and preserving previous knowledge with the replay learning. This enhanced student knowledge is then subsequently harnessed to improve the teacher system, which in turn allows for generating a more powerful student model.
Our contributions are summarized as follows:
\begin{itemize}[leftmargin=*] \vspace{-\topsep}
    \item We explore a new problem of operating teacher-student KD in a non-stationary data stream, which has not been studied well. 
    This direction of research can contribute to bridging the gap between the existing KD techniques and real-world applications. 
    
    \item We introduce \proposed framework, where both the teacher and the student collaboratively evolve along the data stream.
    \proposed facilitates the student's effective adaptation to new data, while also enabling the teacher to fully leverage accumulated knowledge.
    
    \item We validate the effectiveness of \proposed through extensive experiments on real-world datasets.
    Furthermore, we provide thorough analyses to verify the validity of each proposed component.
\end{itemize}\vspace{-\topsep}

\vspace{-0.6cm}
\section{Related Work}
\label{sec:relatedwork}
\smallsection{Knowledge distillation (KD).}
KD serves as a model compression technique that transfers knowledge from a large teacher model to a lightweight student model \cite{KD, FitNet}.
In recent years, the inference costs of recommender systems (RS) have progressively increased, presenting challenges for their practical deployment.
Consequently, KD has gained much research attention to reduce inference costs while maintaining high recommendation performance \cite{BD, TD, hetcomp, zhu2020ensembled, PHR}.
Earlier work \cite{RD} transfers the point-wise importance of top-ranked items by the teacher, and \cite{TD} proposes a topology distillation that transfers relational knowledge from the teacher representation space.
Recent studies \cite{DERRD, DCD, hetcomp, IRRRD, Kang_unbiased} have delved into list-wise distillation to transfer the ranking orders of items directly.
They formulate the distillation as a \textit{ranking matching} problem and train the student to preserve the teacher's permutation.
This approach has shown state-of-the-art performance in various ranking-oriented applications such as recommendation \cite{DERRD, IRRRD, Kang_unbiased} and document retrieval \cite{reddi2021rankdistil, CL-DRD}.
Furthermore, considering the remarkable performance of a massive system consisting of multiple large-scale models, \cite{zhu2020ensembled, hetcomp} introduce distillation methods tailored for compressing knowledge of an ensembled system.
These KD methods have greatly alleviated the huge computational burdens of deploying a large-scale RS, thereby expanding its applicability to various environments.

However, existing KD studies have focused on one-time distillation in static environments, leaving a substantial gap in their application to real-world scenarios with continuously incoming data. 
In this work, we delve into a systematic approach to operating the teacher-student KD in a non-stationary data stream.


\smallsection{Continual learning (CL).}
Continual learning \cite{dualnet, lin2022towards, kirkpatrick2017overcoming}, also known as lifelong learning or incremental learning, is a concept to train a model with a non-stationary data stream.
A major challenge is to strike a balance between plasticity and stability \cite{lin2022towards}, where plasticity refers to the ability to learn new knowledge, and stability focuses on retaining previous knowledge.
Naively updating a model on new data often fails to achieve this balance, resulting in catastrophic forgetting of its previous knowledge \cite{kirkpatrick2017overcoming}.
Two popular CL approaches are regularization \cite{yang2019adaptive, GraphSAIL} and experience replay \cite{shin2017continual, CL_RS_replay}.
The regularization approach typically imposes regularization constraints on the parameter space, discouraging the model parameters from drastically changing from the previously trained ones. 
On the other hand, the replay approach involves reusing historical data, by storing a subset of representative samples \cite{CL_RS_replay} or by using a generative model trained on input distribution \cite{shin2017continual}.

Due to its practical importance, CL has been actively studied for RS.
\cite{GraphSAIL, LWCKD, PIW} have delved into the regularization approach.
\cite{GraphSAIL, LWCKD} have focused on structure-aware regularization for graph-neural network-based models. 
More recently, \cite{PIW} proposes a new personalized weight adjustment to tailor the impact of regularization for each user, pointing out that preserving the same amount of historical information for all users is sub-optimal. 
On the other hand, \cite{zhu2023reloop2, cai2022reloop, CL_RS_replay, mi2020ader} have focused on the replay approach. 
\cite{CL_RS_replay, mi2020ader} select representative data based on interaction frequency, while \cite{cai2022reloop, zhu2023reloop2} have focused on self-correcting learning by using an error memory to store samples that the model previously failed to predict.

Though effective, these CL methods have mostly been studied for updating a \textit{single} model having \textit{sufficient capacity} to learn effectively on its own, which differs significantly from our problem.
Operating the teacher-student KD in a data stream still remains unexplored.

\section{Problem Formulation}
\label{sec:preliminary}

\begin{figure*}[t]
    \centering    
\includegraphics[width=0.80\linewidth]{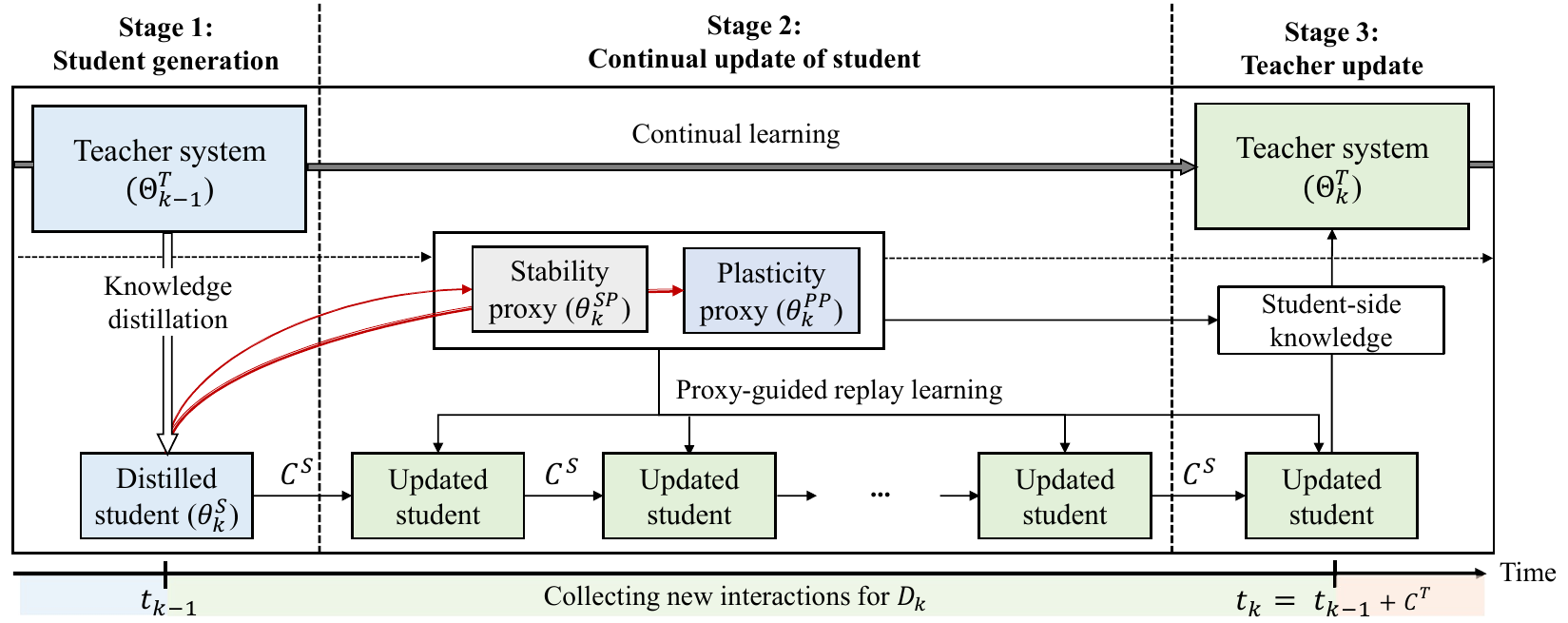}
    \caption{Overview of \proposed framework for $k$-th data block.}
    \label{fig:method}
    \vspace{-0.4cm}
\end{figure*}

\subsection{Concept Definition}

\smallsection{Definition 1} (Teacher-Student Knowledge Distillation).
A massive teacher recommender system \tRS, which typically comprises several large-scale models, achieves high performance through its large capacity.
However, it also incurs high computational costs for inference.
KD is employed to compress \tRS into a lightweight student model $S$ \cite{hetcomp}.
The student model has significantly reduced inference latency, making it well-suited for real-time services and resource-constrained environments.

\smallsection{Definition 2} (Recommendation with Data Stream).
A real-world recommender system operates in a non-stationary data stream where new users, items, and interactions are continuously incoming.
In many practical scenarios, a fixed-size time window of recent data is employed to update the system \cite{GraphSAIL}.
Naturally, the data stream $D$ is viewed as consecutive blocks of interaction data $[D_{1}, D_{2},..., D_{K}]$ with a certain time interval $C$.
$D_k$ corresponds to set of interactions collected from the timestamp $t_{k-1}$ to $t_{k}$, where $t_k = t_{k-1} + C$.
Each block serves as training data for updating the system at each time segment.
At $k$-th block, the system optimizes its performance on $D_{k}$ while leveraging the knowledge obtained from the previous blocks, $D_{1}, ..., D_{k-1}$.
Note that the system only uses the most recent block for training without directly accessing the previous blocks \cite{GraphSAIL, PIW, zhu2023reloop2}.
The performance is evaluated across the entire timeline.

\smallsection{Definition 3} (Teacher/Student update cycle).
An update cycle for a system signifies the frequency at which the system is updated to adapt newly incoming interaction data \cite{mi2020ader}. 
It is often empirically decided considering various factors such as available resources and the amount of collected data.
Let $C^{T}$ and $C^{S}$ denote update cycles for the teacher and the student, respectively.
In practice, $C^{T}$ is much larger than $C^{S}$, as updating the teacher system requires more time and resources compared to the student model.
For instance, one might opt for a weekly update cycle for the teacher ($C^{T} =$ 7 days) while allowing the student to update on a daily basis ($C^{S} =$ 1 day).
It is important to note that we define the data blocks in terms of the teacher system, thus the time interval is defined as $C = C^{T}$.

\subsection{Problem Definition}
We delve into a systematic approach to operate teacher-student KD in a non-stationary data stream.
Our goal is to enable efficient deployment through a lightweight $S$, preserving the high accuracy of \tRS while effectively adapting to continuously incoming data.
This problem has two unique desiderata:
\begin{enumerate}[leftmargin=*]\vspace{-\topsep}
    \item The student should effectively adapt to continuously incoming data by itself during $C^T$.
    That is, while the teacher remains static, the student should cope with evolving preferences as well as new users and items. 
    However, due to its limited capacity, naive updates of the student often lead to ineffective adaptation and a substantial loss of previously acquired knowledge.
    \item The knowledge of teacher and student should be effectively accumulated and leveraged throughout the data stream. 
    The conventional KD approach conducts distillation independently for each time segment.
    This cannot fully leverage the knowledge generated from previous blocks, leading to inefficient training and suboptimal performance.
\end{enumerate}\vspace{-\topsep}
Lastly, it is worth noting that we aim for a model-agnostic solution, enabling service providers to choose their preferred model for both the teacher and the student according to their environments.







\section{METHODOLOGY}
\label{sec:method}

\subsection{Overview}
We introduce a new \textbf{\underline{C}}ontinual \textbf{\underline{C}}ollaborative \textbf{\underline{D}}istillation (\proposed) framework, where both the teacher and the student evolve collaboratively along the non-stationary data stream.
Assume that we have \tRS updated using up to $(k-1)$-th data block.
During the subsequent teacher update cycle (i.e., during collecting $k$-th block for teacher update), \proposed follows three consecutive stages:
\begin{enumerate}[leftmargin=*]\vspace{-\topsep}
    \item \textbf{Student model generation (\cref{sec:s1})}.
    We generate a compact student model $S$ by compressing \tRS through KD. The student model is used for deployment.
    \item \textbf{Continual update of student (\cref{sec:s2})}. 
    The student model is continually updated using non-stationary data with the cycle~$C^S$.
    This process allows recommendations to aptly reflect new users, new items, and up-to-date interactions.
    We introduce new embedding initialization and proxy-guided replay learning strategies to improve the training while preventing the loss of previous knowledge.

    \item \textbf{Teacher system update (\cref{sec:s3})}. When the teacher's next update cycle arrives, \tRS is updated using new interaction data $D_k$ accumulated during the cycle.
    We propose a new strategy for selectively harnessing the knowledge obtained from the student side to further enhance \tRS.
    
\end{enumerate}\vspace{-\topsep}
For each data block, \proposed iterates these three consecutive stages (\cref{sec:s1}-\cref{sec:s3}).
In the subsequent sections, we explain how \proposed operates for the $k$-th data block.
Figure \ref{fig:method} provides an overview~of~\proposed.

\subsection{Stage 1: Student model generation via KD}
\label{sec:s1}
Given the massive teacher system $RS^T\!(\cdot ;\Theta^T_{k-1})$ which was updated up to $(k-1)$-th block, we generate a lightweight student model $S(\cdot ;\theta^S_{k})$ which will be deployed for the current $k$-th block.
As KD is the model-agnostic technique, the student model can be any existing recommendation model that predicts the ranking score for each user-item pair, i.e., $S: \mathcal{U} \times \mathcal{I} \rightarrow \mathbb{R}$, where $\mathcal{U}$ and $\mathcal{I}$ denote the set of users and items, respectively.

In \proposed framework, any off-the-shelf KD technique can be flexibly employed to generate the student model.
In this work, we employ the recent list-wise distillation \cite{hetcomp, DERRD} that trains the student to emulate the item permutation (i.e., ranking orders) predicted by the teacher.
Let $\pi^T_u$ denote the item ranking list for each user $u$ predicted by \tRS.
The list-wise distillation loss is defined as the negative log-likelihood of permutation probability \cite{hetcomp}\footnote{As this KD technique is not our contribution, we provide its details in Appendix \ref{APP:KD}. Note that \proposed framework is not dependent on this specific KD technique.}:
\begin{align}
    \min_{\theta^S} \mathcal{L}_{KD} = -\sum_{u \in \mathcal{U}} \log p(\pi^T_u\mid \theta^S).
    \label{eq:kd}
\end{align}
After KD, the distilled student can achieve low inference latency due to its small size while preserving knowledge of \tRS.

\subsection{Stage 2: Continual update of student with incoming interactions}
\label{sec:s2}
As \tRS cannot be updated during its cycle $C^T$, the student model needs to effectively adapt to continuously incoming data by itself.
Specifically, the student is updated per its own cycle $C^{S}$ ($\ll C^{T}$).

In this section, we explain how we update the student model.
We first present how \proposed learns with new data to provide recommendations reflecting up-to-date interactions (\cref{subsub:s_CF}). 
Then, we introduce a new proxy-guided replay learning to support effective training by preventing the loss of previous knowledge (\cref{subsub:s_proxy}).
Lastly, we summarize the overall training objective (\cref{subsub:s_update}).

\subsubsection{\textbf{Learning with new interactions}}
\label{subsub:s_CF}
We update the student model using new interactions collected for the current data block.
Here, an important challenge arises when dealing with new entities that did not exist in the previous data blocks.
In practical applications, new users and new items continuously emerge. 
However, due to the limited capacity, it is challenging for the student to effectively learn their preference from scratch.

To facilitate learning for these new entities, we introduce a simple yet effective \textit{entity embedding initialization} technique.
Let $G$ denote the user-item bipartite graph, where nodes are users and items, and edges represent their interactions observed within the current data block.
For a given node $i$, we denote $\mathcal{N}^{h}_i$ as the set of $h$-hop neighboring nodes in the graph. 
Recent studies \cite{inmo, inductiveMC_CIKM21} have utilized the 1-hop neighbors to generate embeddings for new entities.
Specifically, the embeddings are initialized by aggregating the directly interacted entity embeddings, i.e., $\mathbf{e}_i = AGG(\{\mathbf{e}_{u}: u \in \mathcal{N}^{1}_i\})$.

However, for new entities, 1-hop interactions are typically highly limited.
To supplement limited interactions, we leverage the 2-hop relations (e.g., items purchased together, users who bought the same item) based on the frequency information from the current data block.
Formally, we identify the set of \textit{prominent} users and items, denoted as $\mathcal{P}_{\mathcal{U}}$ and $\mathcal{P}_{\mathcal{I}}$, which have top interaction frequency in the current block.
These entities reveal the up-to-date prominent trends, such as highly in-demand or trending items, assisting in comprehending newly emerging entities.
We initialize embeddings for new users and items as follows: 
\begin{equation}
\begin{aligned}
    \mathbf{e}_u &= AGG(\{\mathbf{e}_{x}: x \in \mathcal{N}^{1}_u \cup (\mathcal{N}^{2}_u \cap \mathcal{P}_{\mathcal{U}})\}),\\
    \mathbf{e}_i &= AGG(\{\mathbf{e}_{x}: x \in \mathcal{N}^{1}_i \cup (\mathcal{N}^{2}_i \cap \mathcal{P}_{\mathcal{I}})\}).
\end{aligned}
\end{equation}
We use the simple mean pooling for the aggregation function $AGG(\cdot)$.
These initialized embeddings are added to the embedding table of the student model.

Then, the student model is updated with new interactions within the current data block.
We employ the pair-wise ranking loss \cite{BPR}:
\begin{align}\label{BPR}
    \min_{\theta^S}\mathcal{L}_{BPR} = -\sum_{u \in \mathcal{U}} \sum_{i \in \mathcal{N}^{1}_u} \sum_{j \notin \mathcal{N}^{1}_u} \log \sigma(\hat{r}_{u,i}^S - \hat{r}_{u,j}^S),
\end{align}
where $\sigma$ denotes the sigmoid function.
$\hat{r}^S_{u,i} = S(u,i)$ denotes the ranking score predicted by the student.

\subsubsection{\textbf{Proxy-guided replay learning}}
\label{subsub:s_proxy}
Naively updating a model with newly incoming interactions can result in catastrophic forgetting, where the model significantly loses previously acquired knowledge \cite{LWCKD, PIW}.
This largely degrades the overall performance of the model and further hinders the learning of new data.
Furthermore, the student model has a highly limited capacity, which makes it more challenging to update effectively.

As a solution, we introduce a new proxy-guided replay learning, which employs an external memory called \textit{proxy} to assist the student's learning.
The learning theory in neuroscience \cite{CLS} posits that humans do effective learning through two complementary systems: a fast learning system for short-term adaptation to specific experiences, and a slow learning system for gradual acquisition of structured knowledge.
Drawing from the theory, we employ two complementary memories for the student, referred to as \textit{plasticity proxy} focusing on rapid adaptation, and \textit{stability proxy} emphasizing long-term retention of knowledge.
During the student's update, we utilize these proxies to identify forgotten preference knowledge and assist in its preservation by replaying past predictions.

\begin{figure}[t]
    \centering
    \includegraphics[width=0.8\linewidth]{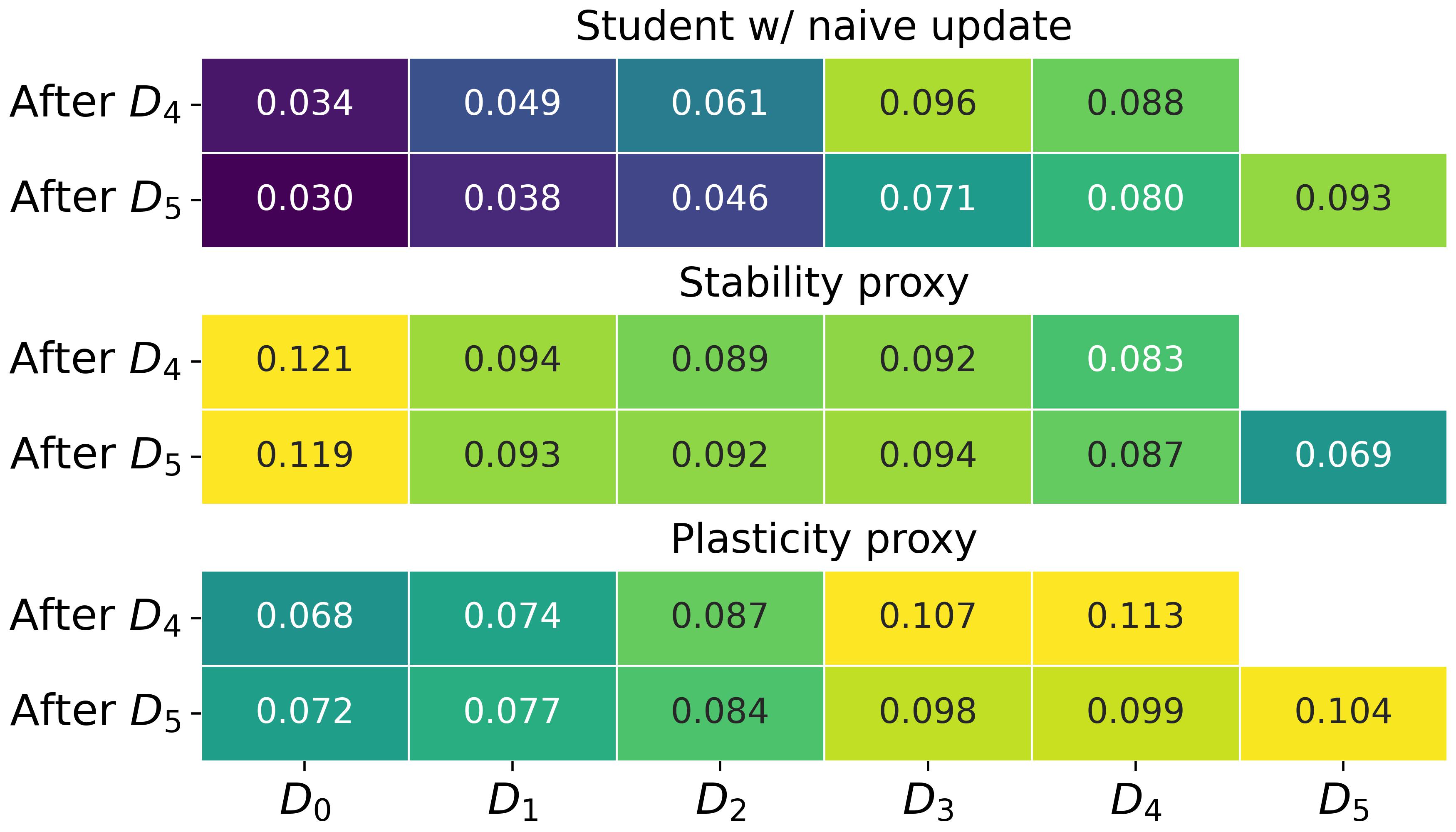}
    \caption{Effects of proxy. After $D_4$ and $D_5$ (y-axis), we assess Recall@20 on test sets from the previous blocks (x-axis). A naive update of the student results in significant catastrophic forgetting. Two proxies effectively accumulate the previous knowledge from complementary perspectives. (Dataset: Yelp)}
    \label{fig:proxy}
\end{figure}

\smallsection{Stability and plasticity proxies.} 
A proxy serves as a memory that accumulates the model knowledge acquired along the data stream.
To efficiently implement the proxy, we employ the \textit{temporal mean} technique \cite{mean_teacher} which accumulates a model's knowledge by the exponential moving average of the model parameters. 
Specifically, we construct the proxy by the temporal mean of the distilled student models generated by KD (\cref{sec:s1}).
We introduce two complementary proxies: the plasticity proxy $S^{PP}$ and the stability proxy $S^{SP}$.
At $k$-th data block, the proxies are updated as follows:
\begin{equation}
\begin{aligned}
\label{eq:proxy}
    \theta^{SP}_{k} = (1 - w_{SP}) \cdot \theta^{SP}_{k-1} + w_{SP} \cdot \theta^{S}_{k}, \\
    \theta^{PP}_{k} = (1 - w_{PP}) \cdot \theta^{PP}_{k-1} + w_{PP} \cdot \theta^{S}_{k}.
\end{aligned}
\end{equation}
$\theta^{S}_{k}$ denotes the parameters of the distilled student model from \cref{sec:s1}.
$\theta^{SP}_{k}$ and $\theta^{PP}_{k}$ denote the parameters for the stability and plasticity proxies that support training on $k$-th block, respectively.
The scalar weights are hyperparameters following the relationship: $0 < w_{SP} \ll w_{PP} < 1$.
Therefore, the stability proxy is slowly updated with the long-term retention of previous knowledge, while the plasticity proxy is rapidly updated with an emphasis on recent trends.
These distinct proxies provide complementary views for the previously acquired knowledge (Figure \ref{fig:proxy}).
Note that proxies are updated once per block, right after KD from the teacher~system.

\smallsection{Replay learning for forgotten knowledge.}
Using the proxies, we identify forgotten preference knowledge and train the student to recover it by replaying past predictions.
A recommendation model produces a ranked list of items for each user. 
In the list, the top-ranked items correspond to the most probable predictions based on the model's knowledge of the user preference.
In this regard, we identify the forgotten knowledge that was previously captured but is not well-reflected in the current model based on rank disparities in the ranking lists;
if items that were previously ranked near the top now hold significantly lower ranks, it may suggest that the model has forgotten knowledge related to them.

Given a model $A$, $\text{rank}_{u,i}^{A}$ denotes the rank of item $i$ in the ranking list of user $u$.
A lower value indicates a higher ranking position, i.e., $\text{rank}_{u,*}^{A}=0$ is the highest rank.
For each item $i$, the rank disparity with respect to another model $B$ is defined as follows:
\begin{equation}
\begin{aligned}
    d^{u}_{i}(A,B) = \exp(\epsilon \cdot (\text{rank}_{u,i}^{A} - \text{rank}_{u,i}^{B})),
\end{aligned}
\end{equation}
where $\epsilon > 0$ is a hyperparameter to control the sharpness of the distribution. 
We utilize the exponential function to put a stronger emphasis on items with large rank disparities.
A high value of $d^{u}_{i}(A,B)$ indicates that item $i$ is ranked significantly higher by model $B$ compared to model $A$.


We create top-$N$ recommendation lists for each user from the student and proxy models, then construct item sets for replay learning based on the rank disparity.
Specifically, we obtain $I_{SP}$ by sampling items from the top-$N$ list of $S^{SP}$, using a probability distribution $p(i) \propto d^{u}_{i}(S,S^{SP})$. 
Similarly, we acquire $I_{PP}$ by sampling items from the top-$N$ list of $S^{PP}$, using a distribution $p(i)\propto d^{u}_{i}(S,S^{PP})$.
Then, we train the student to recover previous knowledge on the identified items by replaying the predictions from~the~proxies:
\begin{equation}
\label{eq:pd}
\begin{aligned}
    \min_{\theta^S} \mathcal{L}_{RE} = -\sum_u (\sum_{i \in I_{SP}} \ell(\hat{r}_{u,i}^S, \hat{r}_{u,i}^{SP}) + \sum_{i \in I_{PP}} \ell(\hat{r}_{u,i}^S, \hat{r}_{u,i}^{PP})).
\end{aligned}
\end{equation}
$\ell(\cdot, \cdot)$ denotes the error function between two predictions.
In this work, we employ the simple binary cross-entropy loss.
It is worth noting that $\mathcal{I}_{SP}$ and $\mathcal{I}_{PP}$ are \textit{dynamically} constructed based on the current state of the student, enabling the effective identification and recovery of forgotten knowledge.
This is an important distinction from the existing replay-based CL methods for RS \cite{CL_RS_replay, zhu2023reloop2, mi2020ader}, which leverage a pre-constructed set of historical data without considering the current state of the student.

\subsubsection{\textbf{The overall objective for student update}}
\label{subsub:s_update}
The student model is updated with the new interaction data per $C^{S}$, assisted by the stability and plasticity proxies.
The overall objective is:
\begin{equation}
\begin{aligned}
\label{eq:s_update}
    \min_{\theta^{S}} \mathcal{L}_{BPR} + \lambda_{RE} \cdot \mathcal{L}_{RE}.
\end{aligned}
\end{equation}
$\lambda_{RE}$ is a hyperparameter controlling the impact of replay learning.

\subsection{Stage 3: Teacher system update}
\label{sec:s3}
Once the teacher update cycle arrives (i.e., the $k$-th block has been completely collected), we update \tRS.
To update a system on the data stream, two types of losses are typically employed \cite{PIW, LWCKD}: (1) collaborative filtering loss to learn new user-item interactions, and (2) continual learning loss to mitigate forgetting of previous system knowledge.
We also employ these two objectives.
Moreover, we introduce an additional objective to fully leverage the knowledge obtained from the student side.

\subsubsection{\textbf{Leveraging the student-side knowledge}}
\label{subsubsec:s_side_knowledge}
We argue that \tRS can be further improved by leveraging knowledge obtained from the student-side in two aspects:
First, the updated student model contains up-to-date knowledge of the current data block, unlike \tRS which has not been updated for a while.
In particular, using the student model, we can obtain potential interactions that are not directly observed in the current block but are likely to be observed in the future.
Second, the proxies, which accumulate knowledge acquired throughout the data stream, serve as valuable knowledge sources to mitigate catastrophic forgetting.
While it is possible to consider utilizing separate proxies for \tRS, given the large size of \tRS, employing the student-side proxies through KD is much more cost-effective in terms of both~space~and~time.

During the teacher training, we selectively leverage the most confident predictions from the student-side models (i.e., the student and two proxies) based on the rank disparity.
Specifically, we identify items ranked near the top by the student-side models but assigned significantly lower rankings by \tRS.
From the top-$N$ list of each student-side model $S^* \in \{S, S^{SP}, S^{PP}\}$, we obtain $I_{S^* \rightarrow T}$ by sampling items using a probability distribution $p(i) \propto d^{u}_{i}(RS^T, S^{*})$.
The student-side knowledge is transferred to the teacher as follows:
\begin{equation}
\label{eq:s_t}
\begin{aligned}
\mathcal{L}_{\text{S} \rightarrow \text{T}} = - \sum_{u} \sum_{S^*} \sum_{i \in I_{S^* \rightarrow T}} \ell(\hat{r}_{u,i}^T, \hat{r}_{u,i}^{S^*}).
\end{aligned}
\end{equation}

\subsubsection{\textbf{The overall objective for teacher update}} 
\tRS is updated with new interaction data, assisted by the student-side knowledge.
\begin{equation}
\label{eq:t_update}
\begin{aligned}
\min_{\Theta^{T}} \ &  \mathcal{L}_{CF} + \lambda_{CL} \cdot \mathcal{L}_{CL} + \lambda_{\text{S} \rightarrow \text{T}} \cdot \mathcal{L}_{\text{S} \rightarrow \text{T}}.
\end{aligned}
\end{equation}
$\Theta^{T}$ denotes the training parameters of \tRS.
$\mathcal{L}_{CF}$ is the original collaborative filtering loss (e.g., binary cross-entropy) used to train \tRS.
$\mathcal{L}_{CL}$ is the continual learning loss. 
\proposed does not require a specific CL technique, and various CL methods for RS can be flexibly employed.
In this work, we use the recently proposed method \cite{PIW}.
$\lambda_{CL}$ and $\lambda_{\text{S} \rightarrow \text{T}}$ are the hyperparameters balancing the losses.
As \tRS evolves during its training, we gradually reduce the impact of the student-side knowledge.
We use a simple annealing schedule, where its impact at $t$-th epoch is controlled as $ \lambda_{\text{S} \rightarrow \text{T}} = \lambda^0_{\text{S} \rightarrow \text{T}}\cdot e^{-t/\tau}$. 
Here, $\lambda^0_{\text{S} \rightarrow \text{T}}$ controls the initial impact, and $\tau$ controls the annealing~speed.
The overall training process is outlined in Algorithm \ref{al1}.

\smallsection{Remarks.}
A key aspect of \proposed involves the collaborative evolution of the teacher and student. 
To elaborate, we enhance the student by effectively adapting to new entities and mitigating the forgetting problem using proxies (\cref{sec:s2}). 
This enhanced student knowledge is then subsequently harnessed to improve the teacher (\cref{sec:s3}), which in turn allows for generating a more powerful student (\cref{sec:s1}).

\begin{algorithm}[t]
\small
\DontPrintSemicolon
\SetKwInOut{Input}{Input}
\SetKwInOut{Output}{Output}
\Input{Data stream $D$, Teacher system $RS^T\!(\cdot ;\Theta^T_{k-1})$, Proxies $S^{SP}(\cdot;\theta^{SP}_{k-1})$, $S^{PP}(\cdot;\theta^{PP}_{k-1})$}
\Output{Updated teacher system $RS^T\!(\cdot ;\Theta^T_{k})$, Updated proxies $S^{SP}(\cdot;\theta^{SP}_{k})$, $S^{PP}(\cdot;\theta^{PP}_{k})$}
\BlankLine
Generate a new distilled student $S$ via KD \Comment{Eq.\ref{eq:kd}} \\
Update the stability and plasticity proxies $S^{SP}$, $S^{PP}$ \Comment{Eq.\ref{eq:proxy}} \\
\For{\textup{every} $C^S$}{
Update the student model $S$ \Comment{Eq.\ref{eq:s_update}} \\
}
Update the teacher system \tRS \Comment{Eq.\ref{eq:t_update}} \\
\BlankLine
\caption{\proposed algorithm for $k$-th data block}
\label{al1}
\end{algorithm}



\section{Experiments}

\subsection{Experimental Setup}
\label{sec:experimentsetup}
We provide details of setup in Appendix \ref{App:setup}.

\subsubsection{\textbf{Datasets.}} 
We use two real-world datasets: Gowalla and Yelp \cite{GraphSAIL, BD}.
To simulate the non-stationary data streams, we split each dataset such that the first 50\% forms the base block ($D_0$), and the remaining 50\% is evenly divided into 5 incremental blocks ($D_1,...,D_5$), according to the temporal timestamp \cite{GraphSAIL, mi2020ader}.
For each incremental block, we randomly divide the interactions of each user into train/validation/test sets in 80\%/10\%/10\% split. 
Table \ref{tab:datablocks} presents the statistics of each block.

\subsubsection{\textbf{Teacher-student KD setup.}}  
As our focus is on operating the teacher-student KD in a data stream, we closely follow the setup of the existing KD studies \cite{hetcomp, RD, DCD}.
For \textbf{teacher}, we construct a massive system by an ensemble of more than five large models \cite{hetcomp}.
We increase the capacity of the teacher system until its performance is no longer improved.
For \textbf{student}, we employ two widely used backbone models: a matrix factorization (MF)-based model \cite{BPR} and a graph neural network (GNN)-based model \cite{he2020lightgcn}.
We set a small embedding size for the student (8 for Yelp, and 16 for Gowalla), considering the teacher size for each dataset \cite{DCD, hetcomp}.
The student update cycle is set as one-tenth of the teacher cycle.
Note that to ensure consistent evaluation for both the teacher and student, all models are assessed on the same test set after training with all interactions within each data block.
Further details on KD setup including teacher configuration, and the number of parameters are provided in Appendix \ref{App:teacher_setup}.

\begin{table}[t]
\caption{Data block statistics of two datasets.}
\centering
\renewcommand{\tabcolsep}{1.1mm}
\resizebox{\columnwidth}{!}{%
\begin{tabular}{cc|c|ccccc}
\hline
\multicolumn{2}{c|}{\textbf{Data Blocks}} & $\mathbf{D_0}$ & $\mathbf{D_1}$ & $\mathbf{D_2}$ & $\mathbf{D_3}$ & $\mathbf{D_4}$ & $\mathbf{D_5}$ \\ \hline\hline
\multicolumn{1}{c|}{} & \textbf{\# of accumulated users} & 19,668 & 21,266 & 24,107 & 26,840 & 29,106 & 29,858 \\
\multicolumn{1}{c|}{} & \textbf{\# of accumulated items} & 39,354 & 39,809 & 40,381 & 40,691 & 40,908 & 40,988 \\
\multicolumn{1}{c|}{} & \textbf{\# of new users} & -- & 1,598 & 2,841 & 2,733 & 2,266 & 752 \\
\multicolumn{1}{c|}{} & \textbf{\# of new items} & -- & 455 & 572 & 310 & 217 & 80 \\
\multicolumn{1}{c|}{\multirow{-5}{*}{\rotatebox[origin=c]{90}{\textbf{Gowalla}} }} & \textbf{\# of interactions} & 513,732 & 62,336 & 112,167 & 123,042 & 126,474 & 89,713 \\ \hline

\multicolumn{1}{c|}{} & \textbf{\# of accumulated users} & 12,248 & 13,367 & 13,896 & 14,399 & 14,761 & 14,950 \\
\multicolumn{1}{c|}{} & \textbf{\# of accumulated items} & 10,822 & 11,345 & 11,634 & 11,970 & 12,197 & 12,261 \\
\multicolumn{1}{c|}{} & \textbf{\# of new users} & -- & 1,119 & 529 & 503 & 362 & 189 \\
\multicolumn{1}{c|}{} & \textbf{\# of new items} & -- & 523 & 289 & 336 & 227 & 64 \\
\multicolumn{1}{c|}{\multirow{-5}{*}{   \rotatebox[origin=c]{90}{\textbf{Yelp}}  }} & \textbf{\# of interactions} & 151,084 & 51,280 & 34,501 & 33,593 & 37,970 & 34,189 \\ \hline
\end{tabular}
}
\label{tab:datablocks}
\vspace{0.1cm}
\end{table}

\subsubsection{\textbf{Evaluation protocol.}}
We closely follow the evaluation protocol of the existing CL methods \cite{do2023continual, GraphSAIL, PIW}.
All models are first trained on $D_0$, then continually updated using the incremental blocks.
Specifically, at $k$-th block, the model is updated using $D_{k}$, while access to past blocks $D_{0}, ..., D_{k-1}$ is forbidden \cite{do2023continual, GraphSAIL, PIW}.

The evaluation of CL focuses on assessing how effectively a model achieves a balance between plasticity and stability.
We employ two CL metrics proposed in \cite{do2023continual}. 
We construct the matrix $A \in \mathbb{R}^{K\times K}$, where each element $a_{ij}$ denotes the recommendation performance on block $j$ after completing training on block $i$.
After training on each $k$-th block, we report two metrics:
\begin{itemize}[leftmargin=*]\vspace{-\topsep}
    \item \textbf{Learning Average (LA)} = $\frac{1}{k} \sum^{k}_{i=1} a_{i,i}$. It assesses how well a model adapts to new blocks, focusing on the plasticity aspect. 
    \item \textbf{Retained Average (RA)} = $\frac{1}{k} \sum^{k}_{i=1} a_{k,i}$. It assesses how well a model retains past knowledge, focusing on the stability aspect.
\end{itemize}\vspace{-\topsep}
A model's overall capability is summarized by the harmonic mean of LA and RA, denoted by \textbf{H-mean} \cite{do2023continual}.
To evaluate the recommendation performance, we employ Recall@20 and NDCG@20 \cite{GraphSAIL, PIW, do2023continual, BUIR}.
We report the results from five independent runs.

\subsubsection{\textbf{Baselines}} 
To evaluate \proposed, which trains both the teacher and student in a data stream, we compare various CL approaches.
\begin{itemize}[leftmargin=*]\vspace{-\topsep}
    \item \textbf{Full-Batch} uses \textit{all the historical data} to train the model from scratch. 
    We report its results solely for reference purposes. 
    \item \textbf{Fine-Tune} is a naive baseline that updates the model using its original loss function on new data.
    \item \textbf{LWC-KD-PIW} \cite{PIW} is the state-of-the-art regularization-based CL method for recommendation.
    It applies the regularization (i.e., LWC-KD \cite{LWCKD}) with personalized weights, adjusting the regularization effects considering the dynamics of each user.
    \item \textbf{ReLoop2} \cite{zhu2023reloop2} is the state-of-the-art replay-based CL method for recommendation.
    It introduces self-correcting learning using an error memory that stores samples that the model failed to~predict.
\end{itemize}\vspace{-\topsep}
\noindent
For the student, we further devise two baselines that represent the best solution that combines the existing KD and CL methods;
along the data stream, we update the teacher using the state-of-the-art CL method (i.e., LWC-KD-PIW).
Then, for each block, we generate a compact student via KD \cite{hetcomp}, which is subsequently updated by either Fine-tune or LWC-KD-PIW, referred to as \textbf{KD + Fine-tune} and \textbf{KD + LWC-KD-PIW}, respectively.
Lastly, our approach is:
\begin{itemize}[leftmargin=*]
    \item \textbf{\proposed} trains the teacher and student collaboratively along the data stream. We use the same KD technique \cite{hetcomp} for stage 1 ($\mathcal{L}_{KD}$ in Eq.\ref{eq:kd}) and LWC-KD-PIW \cite{PIW} for stage 3 ($\mathcal{L}_{CL}$ in Eq.\ref{eq:t_update}).
\end{itemize}


\begin{table*}[ht!]
\caption{The overall performance (Recall@20) comparison. 
* denotes $p < 0.05$ for the paired t-test on \proposed with the best baseline.}
\label{tab:main_table}
\footnotesize
\renewcommand{\arraystretch}{0.85}
\renewcommand{\tabcolsep}{1.2mm}
\centering
\resizebox{\linewidth}{!}{
\begin{tabular}{c|l|ccc|ccc|ccc|ccc|ccc}
\hline
 \multicolumn{2}{c|}{\multirow{2}{*}{\textbf{Gowalla}}} & \multicolumn{3}{c|}{\textbf{After $D_1$}} & \multicolumn{3}{c|}{\textbf{After $D_2$}} & \multicolumn{3}{c|}{\textbf{After $D_3$}} & \multicolumn{3}{c|}{\textbf{After $D_4$}} & \multicolumn{3}{c}{\textbf{After $D_5$}} \\
\multicolumn{2}{c|}{} & \textbf{LA} & \textbf{RA} & \textbf{H-mean} & \textbf{LA} & \textbf{RA} & \textbf{H-mean} & \textbf{LA} & \textbf{RA} & \textbf{H-mean} & \textbf{LA} & \textbf{RA} & \textbf{H-mean} & \textbf{LA} & \textbf{RA} & \textbf{H-mean} \\
\hline\hline
\multirow{6}{*}{\textbf{\begin{tabular}[c]{@{}c@{}}Teacher\end{tabular}}} & Full-Batch  & 0.1758 & 0.1752 & 0.1755 & 0.1760 & 0.1804 & 0.1782 & 0.1739 & 0.1774 & 0.1756 & 0.1753 & 0.1794 & 0.1773 & 0.1767 & 0.1784 & 0.1775 \\ 
 & Fine-Tune & 0.1725 & 0.1544 & 0.1629 & 0.1714 & 0.0957 & 0.1228 & 0.1695 & 0.0828 & 0.1113 & 0.1700 & 0.0717 & 0.1009 & 0.1722 & 0.0577 & 0.0864 \\
 & LWC-KD-PIW & \textbf{0.1826} & 0.1483 & {\ul 0.1637} & \textbf{0.1791} & 0.1141 & 0.1394 & \textbf{0.1772} & 0.1023 & 0.1297 & \textbf{0.1779} & 0.0903 & 0.1198 &  \textbf{0.1797} & 0.0826 & 0.1132 \\
 & ReLoop2 & 0.1706 &
{\ul 0.1570} & 0.1635 & 0.1595 & {\ul 0.1454} & {\ul 0.1521} & 0.1536 & {\ul 0.1243} & {\ul 0.1374} & 0.1512 & {\ul 0.1087} & {\ul 0.1265} & 0.1482 & {\ul 0.0980} & {\ul 0.1180} \\ 
 & \cellcolor{gray!15}\proposed w/ MF student & \cellcolor{gray!15}{\ul 0.1823} & \cellcolor{gray!15}\textbf{0.1623*} & \cellcolor{gray!15}\textbf{0.1717*} & \cellcolor{gray!15}{\ul 0.1784} & \cellcolor{gray!15}\textbf{0.1524*} & \cellcolor{gray!15}\textbf{0.1644*} & \cellcolor{gray!15}0.1763 & \cellcolor{gray!15}\textbf{0.1429*} & \cellcolor{gray!15}\textbf{0.1579*} & \cellcolor{gray!15}{\ul 0.1764} & \cellcolor{gray!15}\textbf{0.1378*} & \cellcolor{gray!15}\textbf{0.1547*} & \cellcolor{gray!15}0.1779 & \cellcolor{gray!15}\textbf{0.1302*} & \cellcolor{gray!15}\textbf{0.1504*} \\
 & \cellcolor{gray!15}\proposed w/ GNN student & \cellcolor{gray!15}0.1809 & \cellcolor{gray!15}0.1619 & \cellcolor{gray!15}0.1709* & \cellcolor{gray!15}{\ul 0.1784} & \cellcolor{gray!15}0.1513* & \cellcolor{gray!15}0.1637* & \cellcolor{gray!15}{\ul 0.1764} & \cellcolor{gray!15}0.1400 & \cellcolor{gray!15}0.1561* & \cellcolor{gray!15}{\ul 0.1764} & \cellcolor{gray!15}0.1333* & \cellcolor{gray!15}0.1519* & \cellcolor{gray!15}{\ul 0.1787} & \cellcolor{gray!15}0.1246* & \cellcolor{gray!15}0.1468* \\
 \hline
\multirow{6}{*}{\textbf{\begin{tabular}[c]{@{}c@{}}Student\\ (MF)\end{tabular}}} & Full-Batch & 0.1303 & 0.1318 & 0.1310 & 0.1296 & 0.1327 & 0.1311 & 0.1293 & 0.1338 & 0.1315 & 0.1310 & 0.1360 & 0.1335 & 0.1324 & 0.1383 & 0.1353 \\
 & LWC-KD-PIW & 0.1261 & 0.1134 & 0.1194 & 0.1281 & 0.0897 & 0.1055 & \textbf{0.1287} & 0.0914 & 0.1069 & \textbf{0.1305} & 0.0789 & 0.0983 & \textbf{0.1342} & 0.0694 & 0.0915 \\
  & ReLoop2 & 0.1101 & 0.1119 & 0.1110 & 0.1066 & 0.1055 & 0.1060 & 0.1023 & 0.0998 & 0.1010 & 0.1008 & 0.0892 & 0.0946 & 0.1003 & 0.0826 & 0.0906 \\
 & KD $+$ Fine-Tune & {\ul 0.1309} & \textbf{0.1457} & {\ul 0.1379} & {\ul 0.1273} & 0.1153 & 0.1210 & 0.1271 & 0.1023 & 0.1134 & 0.1272 & {\ul 0.1031} & {\ul 0.1139} & 0.1292 & 0.0835 & 0.1014 \\
 & KD $+$ LWC-KD-PIW & 0.1292 & {\ul 0.1442} & 0.1363 & 0.1249 & {\ul 0.1235} & {\ul 0.1242} & 0.1247 & {\ul 0.1047} & {\ul 0.1138} & {\ul 0.1284} & 0.0888 & 0.1050 & {\ul 0.1301} & {\ul 0.0842} & {\ul 0.1022} \\
 &\cellcolor{gray!15}\proposed &\cellcolor{gray!15} \textbf{0.1446*} &\cellcolor{gray!15} \textbf{0.1457} &\cellcolor{gray!15} \textbf{0.1451} &\cellcolor{gray!15} \textbf{0.1333} &\cellcolor{gray!15} \textbf{0.1432*} &\cellcolor{gray!15} \textbf{0.1381*} &\cellcolor{gray!15} {\ul0.1280} &\cellcolor{gray!15} \textbf{0.1416*} &\cellcolor{gray!15} \textbf{0.1345*} &\cellcolor{gray!15} 0.1255 &\cellcolor{gray!15} \textbf{0.1363*} &\cellcolor{gray!15} \textbf{0.1307*} &\cellcolor{gray!15} 0.1249 &\cellcolor{gray!15} \textbf{0.1286*} &\cellcolor{gray!15} \textbf{0.1267*} \\
 \hline
\multirow{6}{*}{\textbf{\begin{tabular}[c]{@{}c@{}}Student\\ (GNN)\end{tabular}}} & Full-Batch & 0.1390 & 0.1393 & 0.1391 & 0.1408 & 0.1449 & 0.1428 & 0.1411 & 0.1446 & 0.1428 & 0.1431 & 0.1454 & 0.1442 & 0.1456 & 0.1468 & 0.1462 \\
 & LWC-KD-PIW & 0.1380 & 0.1034 & 0.1182 & 0.1369 & 0.0837 & 0.1039 & 0.1369 & 0.0892 & 0.1080 & 0.1370 & 0.0794 & 0.1005 & 0.1398 & 0.0753 & 0.0979 \\
  & ReLoop2 & 0.1277 & 0.1025 & 0.1137 & 0.1256 & 0.0950 & 0.1082 & 0.1248 & 0.0878 & 0.1031 & 0.1254 & 0.0844 & 0.1009 & 0.1282 & 0.0759 & 0.0953 \\
 & KD $+$ Fine-Tune & {\ul 0.1523} & {\ul 0.1495} & {\ul 0.1509} & 0.1506 & 0.1318 & 0.1406 & 0.1486 & {\ul 0.1115} & {\ul 0.1274} & 0.1496 & {\ul 0.0991} & {\ul 0.1192} & 0.1525 & {\ul 0.0872} & 0.1110 \\
 & KD $+$ LWC-KD-PIW & {\ul 0.1523} & \textbf{0.1496} & {\ul 0.1509} & {\ul 0.1507} & {\ul 0.1323} & {\ul 0.1409} & {\ul 0.1493} & 0.1052 & 0.1234 & {\ul 0.1502} & 0.0966 & 0.1176 & {\ul 0.1534} & 0.0871 & {\ul 0.1111} \\
 &\cellcolor{gray!15}\proposed &\cellcolor{gray!15} \textbf{0.1681*} &\cellcolor{gray!15} 0.1421 &\cellcolor{gray!15} \textbf{0.1540} &\cellcolor{gray!15} \textbf{0.1622} &\cellcolor{gray!15} \textbf{0.1428*} &\cellcolor{gray!15} \textbf{0.1519*} &\cellcolor{gray!15} \textbf{0.1590} &\cellcolor{gray!15} \textbf{0.1376*} &\cellcolor{gray!15} \textbf{0.1475*} &\cellcolor{gray!15} \textbf{0.1584} &\cellcolor{gray!15} \textbf{0.1330*} &\cellcolor{gray!15} \textbf{0.1446*} &\cellcolor{gray!15} \textbf{0.1598} &\cellcolor{gray!15} \textbf{0.1202*} &\cellcolor{gray!15} \textbf{0.1372*} \\

 \hline \hline
\multicolumn{2}{c|}{\multirow{2}{*}{\textbf{Yelp}}} & \multicolumn{3}{c|}{\textbf{After $D_1$}} & \multicolumn{3}{c|}{\textbf{After $D_2$}} & \multicolumn{3}{c|}{\textbf{After $D_3$}} & \multicolumn{3}{c|}{\textbf{After $D_4$}} & \multicolumn{3}{c}{\textbf{After $D_5$}} \\
\multicolumn{2}{c|}{} & \textbf{LA} & \textbf{RA} & \textbf{H-mean} & \textbf{LA} & \textbf{RA} & \textbf{H-mean} & \textbf{LA} & \textbf{RA} & \textbf{H-mean} & \textbf{LA} & \textbf{RA} & \textbf{H-mean} & \textbf{LA} & \textbf{RA} & \textbf{H-mean} \\
\hline\hline
\multirow{6}{*}{\textbf{\begin{tabular}[c]{@{}c@{}}Teacher\end{tabular}}} & Full-Batch & 0.1135 & 0.1145 & 0.1140 & 0.1096 & 0.1164 & 0.1129 & 0.1070 & 0.1159 & 0.1113 & 0.1064 & 0.1141 & 0.1101 & 0.1044 & 0.1151 & 0.1095 \\
                 & Fine-Tune & 0.1210 & 0.1001 & 0.1096 & 0.1122 & {\ul 0.0716} & {\ul 0.0874} & 0.1087 & {\ul 0.0562} & {\ul 0.0741} & 0.1058 & 0.0488 & 0.0668 & 0.1060 & {\ul 0.0548} & {\ul 0.0722} \\
                 & LWC-KD-PIW & {\ul 0.1235} & {\ul 0.1022} & {\ul0.1118} & {\ul 0.1141} & 0.0675 & 0.0848 & {\ul 0.1118} & 0.0552 & 0.0739 & {\ul 0.1083} & 0.0491 & {\ul 0.0676} & {\ul 0.1075} & 0.0499 & 0.0682 \\
                 & ReLoop2 & 0.1051 & 0.0836 & 0.0931 & 0.0949 & 0.0563 & 0.0707 & 0.0927 & 0.0550 & 0.0690 & 0.0898 & {\ul 0.0514} & 0.0654 & 0.0894 & 0.0469 & 0.0615 \\
 & \cellcolor{gray!15}\proposed w/ MF student & \cellcolor{gray!15} \textbf{0.1253} & \cellcolor{gray!15}\textbf{0.1128} &\cellcolor{gray!15} \textbf{0.1187} &\cellcolor{gray!15}\textbf{0.1214*} & \cellcolor{gray!15}\textbf{0.1007*} & \cellcolor{gray!15}\textbf{0.1101*} &\cellcolor{gray!15}\textbf{0.1231*} & \cellcolor{gray!15}\textbf{0.0980*} & \cellcolor{gray!15}\textbf{0.1091*} & \cellcolor{gray!15}\textbf{0.1213*} &\cellcolor{gray!15} 0.0940* & \cellcolor{gray!15}0.1059* & \cellcolor{gray!15}0.1197* & \cellcolor{gray!15}\textbf{0.0921*} & \cellcolor{gray!15}\textbf{0.1041*} \\
 & \cellcolor{gray!15}\proposed w/ GNN student & \cellcolor{gray!15}0.1244 &\cellcolor{gray!15} 0.1079 &\cellcolor{gray!15} 0.1156* &\cellcolor{gray!15} 0.1202 & \cellcolor{gray!15}0.0991* & \cellcolor{gray!15}0.1086* & \cellcolor{gray!15}0.1212 & \cellcolor{gray!15}0.0971* & \cellcolor{gray!15}0.1078* & \cellcolor{gray!15}0.1209 & \cellcolor{gray!15}\textbf{0.0950*} &\cellcolor{gray!15} \textbf{0.1064*} &\cellcolor{gray!15} \textbf{0.1198*} &\cellcolor{gray!15} 0.0914* &\cellcolor{gray!15} 0.1037* \\
\hline
\multirow{6}{*}{\textbf{\begin{tabular}[c]{@{}c@{}}Student\\ (MF)\end{tabular}}} & Full-Batch & 0.0805 & 0.0805 & 0.0805 & 0.0766 & 0.0785 & 0.0775 & 0.0746 & 0.0755 & 0.0750 & 0.0723 & 0.0764 & 0.0743 & 0.0705 & 0.0767 & 0.0735 \\
 & LWC-KD-PIW & 0.0811 & 0.0795 & 0.0803 & 0.0770 & 0.0621 & 0.0688 & 0.0743 & 0.0591 & 0.0658 & 0.0725 & 0.0553 & 0.0627 & 0.0719 & 0.0516 & 0.0601 \\
  & ReLoop2 & 0.0796 & 0.0796 & 0.0796 & 0.0775 & 0.0778 & 0.0776 & 0.0734 & 0.0748 & 0.0741 & 0.0699 & 0.0690 & 0.0694 & 0.0671 & 0.0643 & 0.0657 \\
 & KD $+$ Fine-Tune & {\ul 0.0882} & \textbf{0.1058} & {\ul 0.0962} & 0.0858 & 0.0962 & 0.0907 & {\ul 0.0846} & 0.0879 & 0.0862 & {\ul 0.0843} & 0.0822 & 0.0832 & 0.0806 & {\ul 0.0757} & {\ul 0.0781} \\
 & KD $+$ LWC-KD-PIW & 0.0863 & 0.1037 & 0.0942 & {\ul 0.0867} & {\ul 0.0995} & {\ul 0.0927} & 0.0843 & {\ul 0.0885} & {\ul 0.0863} & 0.0842 & {\ul 0.0824} & {\ul 0.0833} & {\ul 0.0833} & 0.0735 & {\ul 0.0781} \\
 &\cellcolor{gray!15}\proposed &\cellcolor{gray!15} \textbf{0.1037*} &\cellcolor{gray!15} {\ul 0.1043} &\cellcolor{gray!15} \textbf{0.1040*} &\cellcolor{gray!15} \textbf{0.0978*} &\cellcolor{gray!15} \textbf{0.1004} &\cellcolor{gray!15} \textbf{0.0991*} &\cellcolor{gray!15} \textbf{0.0923*} &\cellcolor{gray!15} \textbf{0.0902} &\cellcolor{gray!15} \textbf{0.0912*} &\cellcolor{gray!15} \textbf{0.0901*} &\cellcolor{gray!15} \textbf{0.0873} &\cellcolor{gray!15} \textbf{0.0887*} &\cellcolor{gray!15} \textbf{0.0885*} &\cellcolor{gray!15} \textbf{0.0870*} &\cellcolor{gray!15} \textbf{0.0877*} \\
 \hline
\multirow{6}{*}{\textbf{\begin{tabular}[c]{@{}c@{}}Student\\ (GNN)\end{tabular}}} & Full-Batch & 0.0873 & 0.0894 & 0.0883 & 0.0840 & 0.0874 & 0.0857 & 0.0824 & 0.0884 & 0.0853 & 0.0817 & 0.0898 & 0.0856 & 0.0799 & 0.0890 & 0.0842 \\
 & LWC-KD-PIW & 0.0937 & 0.0726 & 0.0818 & 0.0878 & 0.0495 & 0.0633 & 0.0876 & 0.0442 & 0.0588 & 0.0856 & 0.0383 & 0.0529 & 0.0856 & 0.0383 & 0.0529 \\
  & ReLoop2 & 0.0825 & 0.0726 & 0.0772 & 0.0769 & 0.0550 & 0.0641 & 0.0765 & 0.0494 & 0.0600 & 0.0752 & 0.0528 & 0.0620 & 0.0762 & 0.0510 & 0.0611 \\
 & KD $+$ Fine-Tune & {\ul 0.0997} & {\ul 0.1022} & {\ul 0.1009} & {\ul 0.0992} & {\ul 0.0940} & {\ul 0.0965} & {\ul 0.0970} & 0.0869 & {\ul 0.0917} & {\ul 0.0953} & {\ul 0.0845} & {\ul 0.0896} & {\ul 0.0974} & 0.0808 & {\ul 0.0883} \\
 & KD $+$ LWC-KD-PIW & 0.0986 & 0.1007 & 0.0996 & 0.0974 & 0.0928 & 0.0950 & 0.0954 & {\ul 0.0874} & 0.0912 & 0.0936 & 0.0843 & 0.0887 & 0.0958 & {\ul 0.0809} & 0.0877 \\
 & \cellcolor{gray!15}\proposed & \cellcolor{gray!15}\textbf{0.1136*} & \cellcolor{gray!15}\textbf{0.1054*} & \cellcolor{gray!15}\textbf{0.1093*} & \cellcolor{gray!15}\textbf{0.1084*} & \cellcolor{gray!15}\textbf{0.0982} & \cellcolor{gray!15}\textbf{0.1030*} & \cellcolor{gray!15}\textbf{0.1097*} & \cellcolor{gray!15}\textbf{0.0930} & \cellcolor{gray!15}\textbf{0.1007*} & \cellcolor{gray!15}\textbf{0.1090*} & \cellcolor{gray!15}\textbf{0.0913} & \cellcolor{gray!15}\textbf{0.0994*} & \cellcolor{gray!15}\textbf{0.1100*} & \cellcolor{gray!15}\textbf{0.0878} & \cellcolor{gray!15}\textbf{0.0977*}\\ 
 \hline
\end{tabular}
}
\vspace{-0.3cm}
\end{table*}


\begin{table*}[ht!]
\caption{The overall performance (NDCG@20) comparison. * denotes $p < 0.05$ for the paired t-test on \proposed with the best baseline.}
\label{tab:main_table_ndcg}
\footnotesize
\renewcommand{\arraystretch}{0.85}
\renewcommand{\tabcolsep}{1.2mm}
\centering
\resizebox{\linewidth}{!}{
\begin{tabular}{cl|ccc|ccc|ccc|ccc|ccc}
\hline
\multicolumn{2}{c|}{\multirow{2}{*}{\textbf{Gowalla}}}                                                                               & \multicolumn{3}{c|}{\textbf{After $D_1$}}                 & \multicolumn{3}{c|}{\textbf{After $D_2$}}                 & \multicolumn{3}{c|}{\textbf{After $D_3$}}                 & \multicolumn{3}{c|}{\textbf{After $D_4$}}                 & \multicolumn{3}{c}{\textbf{After $D_5$}}                  \\
\multicolumn{2}{c|}{}                                                                                                                & \textbf{LA}      & \textbf{RA}      & \textbf{H-mean}  & \textbf{LA}      & \textbf{RA}      & \textbf{H-mean}  & \textbf{LA}      & \textbf{RA}      & \textbf{H-mean}  & \textbf{LA}      & \textbf{RA}      & \textbf{H-mean}  & \textbf{LA}      & \textbf{RA}      & \textbf{H-mean}  \\ \hline\hline
\multicolumn{1}{c|}{\multirow{6}{*}{\textbf{Teacher}}}                                                 & Full-Batch                  & 0.0897           & 0.0893           & 0.0895           & 0.0879           & 0.0900           & 0.0889           & 0.0864           & 0.0874           & 0.0869           & 0.0863           & 0.0870           & 0.0866           & 0.0864           & 0.0860           & 0.0862           \\
\multicolumn{1}{c|}{}                                                                                  & Fine-Tune                   & 0.0858           & 0.0752           & 0.0802           & 0.0848           & 0.0457           & 0.0594           & 0.0837           & 0.0380           & 0.0523           & 0.0838           & 0.0329           & 0.0472           & 0.0852           & 0.0268           & 0.0408           \\
\multicolumn{1}{c|}{}                                                                                  & LWC-KD-PIW                  & \textbf{0.0919}  & 0.0725           & {\ul 0.0811}     & \textbf{0.0889}  & 0.0543           & 0.0674           & \textbf{0.0875}  & 0.0477           & 0.0617           & \textbf{0.0876}  & 0.0426           & 0.0573           & \textbf{0.0885}  & 0.0390           & 0.0541           \\
\multicolumn{1}{c|}{}                                                                                  & ReLoop2                     & 0.0853           & {\ul 0.0769}     & 0.0809           & 0.0777           & {\ul 0.0697}     & {\ul 0.0735}     & 0.0741           & {\ul 0.0589}     & {\ul 0.0656}     & 0.0721           & {\ul 0.0506}     & {\ul 0.0595}     & 0.0702           & {\ul 0.0455}     & {\ul 0.0552}     \\
\multicolumn{1}{c|}{}                                                                                  &\cellcolor{gray!15} CCD w/ MF student           &\cellcolor{gray!15} {\ul 0.0915}     &\cellcolor{gray!15} \textbf{0.0804}  &\cellcolor{gray!15} \textbf{0.0856*} &\cellcolor{gray!15} {\ul 0.0877}     &\cellcolor{gray!15} \textbf{0.0731*} &\cellcolor{gray!15} \textbf{0.0797*} &\cellcolor{gray!15} {\ul 0.0857}     &\cellcolor{gray!15} \textbf{0.0681*} &\cellcolor{gray!15} \textbf{0.0759*} &\cellcolor{gray!15} {\ul 0.0851}     &\cellcolor{gray!15} \textbf{0.0651*} &\cellcolor{gray!15} \textbf{0.0738*} &\cellcolor{gray!15} {\ul 0.0853}     &\cellcolor{gray!15} \textbf{0.0606*} &\cellcolor{gray!15} \textbf{0.0709*} \\
\multicolumn{1}{c|}{}                                                                                  &\cellcolor{gray!15} CCD w/ GNN student          &\cellcolor{gray!15} 0.0903           &\cellcolor{gray!15} 0.0794           &\cellcolor{gray!15} 0.0845*          &\cellcolor{gray!15} 0.0871           &\cellcolor{gray!15} 0.0718           &\cellcolor{gray!15} 0.0787*          &\cellcolor{gray!15} 0.0854           &\cellcolor{gray!15} 0.0660*           &\cellcolor{gray!15} 0.0745*          &\cellcolor{gray!15} 0.0848           &\cellcolor{gray!15} 0.0628*          &\cellcolor{gray!15} 0.0722*          &\cellcolor{gray!15} 0.0852           &\cellcolor{gray!15} 0.0584*          &\cellcolor{gray!15} 0.0693*          \\ \hline
\multicolumn{1}{c|}{\multirow{6}{*}{\textbf{\begin{tabular}[c]{@{}c@{}}Student\\ (MF)\end{tabular}}}}  & Full-Batch                  & 0.0664           & 0.0665           & 0.0664           & 0.0638           & 0.0645           & 0.0641           & 0.0635           & 0.0649           & 0.0642           & 0.0633           & 0.0652           & 0.0642           & 0.0634           & 0.0653           & 0.0643           \\
\multicolumn{1}{c|}{}                                                                                  & LWC-KD-PIW                  & 0.0620           & 0.0554           & 0.0585           & {\ul 0.0619}     & 0.0426           & 0.0505           & {\ul 0.0616}     & 0.0427           & 0.0504           & \textbf{0.0620}  & 0.0368           & 0.0462           & \textbf{0.0641}  & 0.0328           & 0.0434           \\
\multicolumn{1}{c|}{}                                                                                  & ReLoop2                     & 0.0550           & 0.0554           & 0.0552           & 0.0515           & 0.0506           & 0.0510           & 0.0491           & 0.0473           & 0.0482           & 0.0478           & 0.0418           & 0.0446           & 0.0472           & 0.0382           & 0.0422           \\
\multicolumn{1}{c|}{}                                                                                  & KD + Fine-Tune              & {\ul 0.0643}     & {\ul 0.0731}     & {\ul 0.0684}     & 0.0610           & 0.0549           & 0.0578           & 0.0607           & 0.0481           & {\ul 0.0537}     & 0.0603           & {\ul 0.0485}     & {\ul 0.0538}     & 0.0611           & 0.0389           & 0.0475           \\
\multicolumn{1}{c|}{}                                                                                  & KD + LWC-KD-PIW             & 0.0641           & 0.0726           & 0.0681           & 0.0607           & {\ul 0.0601}     & {\ul 0.0604}     & 0.0601           & {\ul 0.0486}     & {\ul 0.0537}     & {\ul 0.0612}     & 0.0419           & 0.0497           & {\ul 0.0616}     & {\ul 0.0392}     & {\ul 0.0479}     \\
\multicolumn{1}{c|}{}                                                                                  &\cellcolor{gray!15} CCD                        &\cellcolor{gray!15} \textbf{0.0733*} &\cellcolor{gray!15} \textbf{0.0738}  &\cellcolor{gray!15} \textbf{0.0735*} &\cellcolor{gray!15} \textbf{0.0653*} &\cellcolor{gray!15} \textbf{0.0683*} &\cellcolor{gray!15} \textbf{0.0668*} &\cellcolor{gray!15} \textbf{0.0622}  &\cellcolor{gray!15} \textbf{0.0674*} &\cellcolor{gray!15} \textbf{0.0647*} &\cellcolor{gray!15} 0.0602           &\cellcolor{gray!15} \textbf{0.0643*} &\cellcolor{gray!15} \textbf{0.0622*} &\cellcolor{gray!15} 0.0591           &\cellcolor{gray!15} \textbf{0.0600*}   &\cellcolor{gray!15} \textbf{0.0595*} \\ \hline
\multicolumn{1}{c|}{\multirow{6}{*}{\textbf{\begin{tabular}[c]{@{}c@{}}Student\\ (GNN)\end{tabular}}}} & Full-Batch                  & 0.0709           & 0.0707           & 0.0708           & 0.0700           & 0.0719           & 0.0709           & 0.0696           & 0.0706           & 0.0701           & 0.0696           & 0.0702           & 0.0699           & 0.0701           & 0.0701           & 0.0701           \\
\multicolumn{1}{c|}{}                                                                                  & LWC-KD-PIW                  & 0.0692           & 0.0496           & 0.0578           & 0.0682           & 0.0398           & 0.0503           & 0.0672           & 0.0405           & 0.0505           & 0.0665           & 0.0384           & 0.0487           & 0.0682           & 0.0278           & 0.0395           \\
\multicolumn{1}{c|}{}                                                                                  & ReLoop2                     & 0.0550           & 0.0554           & 0.0552           & 0.0515           & 0.0506           & 0.0510           & 0.0491           & 0.0473           & 0.0482           & 0.0478           & 0.0418           & 0.0446           & 0.0472           & 0.0382           & 0.0422           \\
\multicolumn{1}{c|}{}                                                                                  & KD + Fine-Tune              & \textbf{0.0755}  & {\ul 0.0697}     & \textbf{0.0725}  & \textbf{0.0731}  & {\ul 0.0591}     & {\ul 0.0654}     & \textbf{0.0717}  & 0.0477           & 0.0573           & \textbf{0.0714}  & 0.0448           & 0.0551           & \textbf{0.0732}  & {\ul 0.0385}     & {\ul 0.0505}     \\
\multicolumn{1}{c|}{}                                                                                  & KD + LWC-KD-PIW             & {\ul 0.0751}     & 0.0695           & {\ul 0.0722}     & {\ul 0.0729}     & 0.0589           & 0.0652           & {\ul 0.0715}     & {\ul 0.0491}     & {\ul 0.0582}     & {\ul 0.0712}     & {\ul 0.0454}     & {\ul 0.0554}     & {\ul 0.0730}     & 0.0381           & 0.0501           \\
\multicolumn{1}{c|}{}                                                                                  &\cellcolor{gray!15} CCD                        &\cellcolor{gray!15} 0.0716           &\cellcolor{gray!15} \textbf{0.0715}  &\cellcolor{gray!15} 0.0715           &\cellcolor{gray!15} 0.0707           &\cellcolor{gray!15} \textbf{0.0674*} &\cellcolor{gray!15} \textbf{0.0690*}  &\cellcolor{gray!15} 0.0703           &\cellcolor{gray!15} \textbf{0.0650*}  &\cellcolor{gray!15} \textbf{0.0675*} &\cellcolor{gray!15} 0.0704           &\cellcolor{gray!15} \textbf{0.0620*}  &\cellcolor{gray!15} \textbf{0.0659*} &\cellcolor{gray!15} 0.0712           &\cellcolor{gray!15} \textbf{0.0557*} &\cellcolor{gray!15} \textbf{0.0625*} \\ \hline\hline
\multicolumn{2}{c|}{\multirow{2}{*}{\textbf{Yelp}}}                                                                                  & \multicolumn{3}{c|}{\textbf{After $D_1$}}                 & \multicolumn{3}{c|}{\textbf{After $D_2$}}                 & \multicolumn{3}{c|}{\textbf{After $D_3$}}                 & \multicolumn{3}{c|}{\textbf{After $D_4$}}                 & \multicolumn{3}{c}{\textbf{After $D_5$}}                  \\
\multicolumn{2}{c|}{}                                                                                                                & \textbf{LA}      & \textbf{RA}      & \textbf{H-mean}  & \textbf{LA}      & \textbf{RA}      & \textbf{H-mean}  & \textbf{LA}      & \textbf{RA}      & \textbf{H-mean}  & \textbf{LA}      & \textbf{RA}      & \textbf{H-mean}  & \textbf{LA}      & \textbf{RA}      & \textbf{H-mean}  \\ \hline\hline
\multicolumn{1}{c|}{\multirow{6}{*}{\textbf{Teacher}}}                                                 & Full-Batch                  & 0.0468           & 0.0484           & 0.0476           & 0.0444           & 0.0474           & 0.0459           & 0.0433           & 0.0468           & 0.0450           & 0.0426           & 0.0473           & 0.0448           & 0.0413           & 0.0449           & 0.0430           \\
\multicolumn{1}{c|}{}                                                                                  & Fine-Tune                   & 0.0502           & {\ul 0.0403}     & {\ul 0.0447}     & {\ul 0.0462}     & {\ul 0.0304}     & {\ul 0.0367}     & {\ul 0.0451}     & {\ul 0.0263}     & {\ul 0.0332}     & {\ul 0.0438}     & 0.0215           & {\ul 0.0288}     & {\ul 0.0435}     & {\ul 0.0209}     & {\ul 0.0282}     \\
\multicolumn{1}{c|}{}                                                                                  & LWC-KD-PIW                  & {\ul 0.0505}     & 0.0379           & 0.0433           & 0.0461           & 0.0259           & 0.0332           & 0.0447           & 0.0216           & 0.0291           & 0.0431           & 0.0175           & 0.0249           & 0.0429           & 0.0169           & 0.0242           \\
\multicolumn{1}{c|}{}                                                                                  & ReLoop2                     & 0.0449           & 0.0336           & 0.0384           & 0.0416           & 0.0257           & 0.0318           & 0.0400           & 0.0228           & 0.0290           & 0.0386           & {\ul 0.0219}     & 0.0279           & 0.0383           & 0.0205           & 0.0267           \\
\multicolumn{1}{c|}{}                                                                                  &\cellcolor{gray!15} CCD w/ MF student  &\cellcolor{gray!15} \textbf{0.0518}  &\cellcolor{gray!15} \textbf{0.0454*} &\cellcolor{gray!15} \textbf{0.0484*} &\cellcolor{gray!15} \textbf{0.0499*} &\cellcolor{gray!15} \textbf{0.0396*} &\cellcolor{gray!15} \textbf{0.0442*} &\cellcolor{gray!15} \textbf{0.0506*} &\cellcolor{gray!15} \textbf{0.0388*} &\cellcolor{gray!15} \textbf{0.0439*} &\cellcolor{gray!15} \textbf{0.0499*} &\cellcolor{gray!15} \textbf{0.0369*} &\cellcolor{gray!15} \textbf{0.0424*} &\cellcolor{gray!15} \textbf{0.0495*} &\cellcolor{gray!15} \textbf{0.0356*} &\cellcolor{gray!15} \textbf{0.0414*} \\
\multicolumn{1}{c|}{}                                                                                  &\cellcolor{gray!15} CCD w/ GNN student &\cellcolor{gray!15} 0.0496           &\cellcolor{gray!15} 0.0414           &\cellcolor{gray!15} 0.0451           &\cellcolor{gray!15} 0.0471           &\cellcolor{gray!15} 0.0384*          &\cellcolor{gray!15} 0.0423*          &\cellcolor{gray!15} 0.0465           &\cellcolor{gray!15} 0.0372*          &\cellcolor{gray!15} 0.0413*          &\cellcolor{gray!15} 0.0460*           &\cellcolor{gray!15} 0.0351*          &\cellcolor{gray!15} 0.0398*          &\cellcolor{gray!15} 0.0458           &\cellcolor{gray!15} 0.0334*          &\cellcolor{gray!15} 0.0386*          \\ \hline
\multicolumn{1}{c|}{\multirow{6}{*}{\textbf{\begin{tabular}[c]{@{}c@{}}Student\\ (MF)\end{tabular}}}}  & Full-Batch                  & 0.0334           & 0.0332           & 0.0333           & 0.0307           & 0.0310           & 0.0308           & 0.0297           & 0.0299           & 0.0298           & 0.0287           & 0.0302           & 0.0294           & 0.0280           & 0.0304           & 0.0292           \\
\multicolumn{1}{c|}{}                                                                                  & LWC-KD-PIW                  & 0.0333           & 0.0325           & 0.0329           & 0.0313           & 0.0246           & 0.0275           & 0.0298           & 0.0236           & 0.0263           & 0.0290           & 0.0222           & 0.0251           & 0.0285           & 0.0204           & 0.0238           \\
\multicolumn{1}{c|}{}                                                                                  & ReLoop2                     & 0.0327           & 0.0327           & 0.0327           & 0.0310           & 0.0314           & 0.0312           & 0.0292           & 0.0297           & 0.0294           & 0.0279           & 0.0279           & 0.0279           & 0.0266           & 0.0253           & 0.0259           \\
\multicolumn{1}{c|}{}                                                                                  & KD + Fine-Tune              & {\ul 0.0361}     & \textbf{0.0435}  & {\ul 0.0395}     & {\ul 0.0346}     & {\ul 0.0389}     & {\ul 0.0366}     & {\ul 0.0339}     & \textbf{0.0359}  & {\ul 0.0349}     & {\ul 0.0338}     & {\ul 0.0334}     & {\ul 0.0336}     & 0.0317           & {\ul 0.0303}     & {\ul 0.0310}     \\
\multicolumn{1}{c|}{}                                                                                  & KD + LWC-KD-PIW             & 0.0358           & 0.0422           & 0.0387           & 0.0336           & 0.0365           & 0.0350           & 0.0336           & 0.0343           & 0.0339           & 0.0333           & 0.0290           & 0.0310           & {\ul 0.0332}     & 0.0264           & 0.0294           \\
\multicolumn{1}{c|}{}                                                                                  &\cellcolor{gray!15} CCD               &\cellcolor{gray!15} \textbf{0.0424*} &\cellcolor{gray!15} {\ul 0.0428}     &\cellcolor{gray!15} \textbf{0.0426*} &\cellcolor{gray!15} \textbf{0.0392*} &\cellcolor{gray!15} \textbf{0.0399}  &\cellcolor{gray!15} \textbf{0.0395*} &\cellcolor{gray!15} \textbf{0.0370*}  &\cellcolor{gray!15} {\ul 0.0357}     &\cellcolor{gray!15} \textbf{0.0363*} &\cellcolor{gray!15} \textbf{0.0358*} &\cellcolor{gray!15} \textbf{0.0342}  &\cellcolor{gray!15} \textbf{0.0350*}  &\cellcolor{gray!15} \textbf{0.0349}  &\cellcolor{gray!15} \textbf{0.0332*} &\cellcolor{gray!15} \textbf{0.0340*}  \\ \hline
\multicolumn{1}{c|}{\multirow{6}{*}{\textbf{\begin{tabular}[c]{@{}c@{}}Student\\ (GNN)\end{tabular}}}} & Full-Batch                  & 0.0363           & 0.0370           & 0.0366           & 0.0339           & 0.0355           & 0.0347           & 0.0331           & 0.0357           & 0.0344           & 0.0326           & 0.0356           & 0.0340           & 0.0320           & 0.0353           & 0.0336           \\
\multicolumn{1}{c|}{}                                                                                  & LWC-KD-PIW                  & 0.0397           & 0.0293           & 0.0337           & 0.0367           & 0.0195           & 0.0255           & 0.0365           & 0.0177           & 0.0238           & 0.0354           & 0.0147           & 0.0208           & 0.0357           & 0.0153           & 0.0214           \\
\multicolumn{1}{c|}{}                                                                                  & ReLoop2                     & 0.0344           & 0.0291           & 0.0315           & 0.0319           & 0.0221           & 0.0261           & 0.0312           & 0.0191           & 0.0237           & 0.0306           & 0.0207           & 0.0247           & 0.0315           & 0.0200           & 0.0245           \\
\multicolumn{1}{c|}{}                                                                                  & KD + Fine-Tune              & 0.0406           & \textbf{0.0416}  & {\ul 0.0411}           & 0.0398           & {\ul 0.0370}     & {\ul 0.0383}     & 0.0386           & 0.0342           & 0.0363           & 0.0376           & {\ul 0.0329}     & {\ul 0.0351}     & 0.0385           & {\ul 0.0310}     & {\ul 0.0343}     \\
\multicolumn{1}{c|}{}                                                                                  & KD + LWC-KD-PIW             & {\ul 0.0410}     & {\ul 0.0414}     & \textbf{0.0412}  & {\ul 0.0400}     & 0.0367           & {\ul 0.0383}     & {\ul 0.0391}     & {\ul 0.0348}     & {\ul 0.0368}     & {\ul 0.0379}     & 0.0324           & 0.0349           & {\ul 0.0387}     & 0.0307           & 0.0342           \\
\multicolumn{1}{c|}{}                                                                                  &\cellcolor{gray!15} CCD               &\cellcolor{gray!15} \textbf{0.0411}  &\cellcolor{gray!15} 0.0412           &\cellcolor{gray!15} {\ul 0.0411}           &\cellcolor{gray!15} \textbf{0.0401}  &\cellcolor{gray!15} \textbf{0.0392}  &\cellcolor{gray!15} \textbf{0.0396*} &\cellcolor{gray!15} \textbf{0.0411*} &\cellcolor{gray!15} \textbf{0.0366*} &\cellcolor{gray!15} \textbf{0.0387*} &\cellcolor{gray!15} \textbf{0.0407*} &\cellcolor{gray!15} \textbf{0.0355*} &\cellcolor{gray!15} \textbf{0.0379*} &\cellcolor{gray!15} \textbf{0.0415*} &\cellcolor{gray!15} \textbf{0.0339*} &\cellcolor{gray!15} \textbf{0.0373*} \\ \hline
\end{tabular}
}\vspace{-0.35cm}
\end{table*}

\subsubsection{\textbf{Implementation Details.}}
We utilize PyTorch with CUDA from RTX A6000 and AMD EPYC 7313 CPU.
All hyperparameters are tuned via grid search on the validation set. 
The learning rate and $L_2$ regularization for the Adam optimizer are chosen from \{0.001, 0.005, 0.01\} and \{0.0001, 0.0005, 0.001\}, respectively. 
For hyperparameters related to the list-wise KD, we follow the setup from \cite{hetcomp}. 
The impacts of continual learning $\lambda_{CL}$ is chosen from \{0.001, 0.01, 0.1\}.
For \proposed, $w_{SP}$ and $w_{PP}$ for updating stability and plasticity proxies are chosen from \{0.0, 0.1\} and \{0.9, 1.0\}, respectively. 
The number of item samples $I_{PP}$, $I_{SP}$, $I_{S^*\rightarrow T}$ is chosen from \{1, 3, 5\}. 
The remaining hyperparameters of \proposed are set to their default values.
The number of prominent users $\mathcal{P}_{\mathcal{U}}$ and items $\mathcal{P}_{\mathcal{I}}$ are set to 20. 
The rank disparity-based sampling is conducted from the top-50 list of each model (i.e., $N=50$).
We set $\epsilon = 10^{-3}$, $\tau=5$, $\lambda_{RE}=0.5$, and $\lambda^0_{\text{S} \rightarrow \text{T}}=0.5$.
For baseline-specific hyperparameters, we follow the search ranges provided in the original papers.

\subsection{Performance Comparison}
\label{sec:result}

\subsubsection{\textbf{Main results.}}
\begin{figure}[t!]
  \includegraphics[width=0.49\columnwidth]{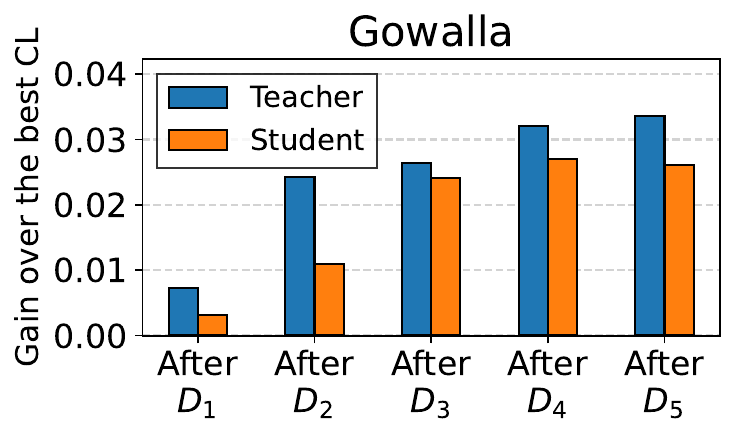}
    \hfill
    \includegraphics[width=0.49\columnwidth]{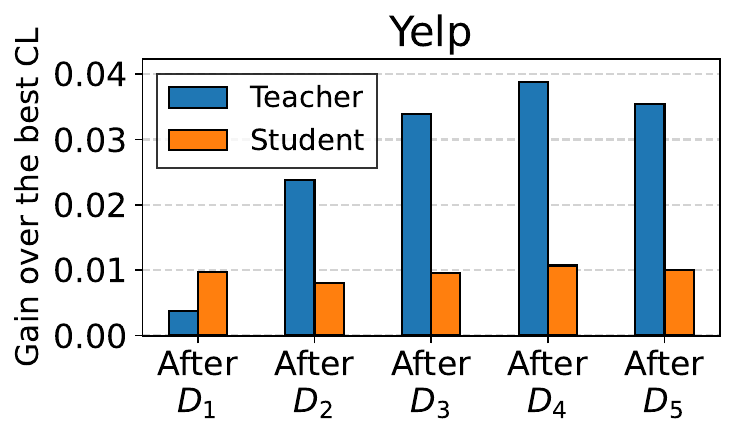}
  \caption{H-mean gain of \proposed over the best CL competitor.}
  \label{fig:gap_images}
\end{figure}

Table \ref{tab:main_table} and Table \ref{tab:main_table_ndcg} show the overall performances of the teacher systems and student models optimized by each compared method.
Overall, \proposed performs better than all baselines in terms of both adapting to new data (LA) and retaining previous knowledge (RA), achieving a good balance between them (H-mean).
Also, \proposed consistently improves both the teacher and student, with the improvements often gradually increasing throughout the data stream (Figure \ref{fig:gap_images}).
We analyze the results from various perspectives:

\smallsection{Teacher-side.}
Teacher systems trained with \proposed consistently achieve higher H-mean across all data blocks compared to state-of-the-art CL methods, including both regularization-based (i.e., LWC-KD-PIW) and replay-based (i.e., ReLoop2) methods.
Unlike the CL methods that update the teacher by itself, \proposed additionally leverages the student-side knowledge.
As discussed in \cref{subsubsec:s_side_knowledge}, the student-side models provide up-to-date knowledge from the updated student as well as historical knowledge accumulated via two complementary proxies.
This can aid the teacher in adapting to new data more effectively while retaining past knowledge.
Interestingly, we observe that the forgetting phenomenon is exacerbated through the ensemble for the teacher.
We conjecture that this amplification could stem from the combined effect of individual model forgetting within the ensemble, leaving further investigation~for~future~study.

\smallsection{Student-side.}
Student models trained with \proposed consistently achieve higher H-mean compared to both state-of-the-art CL methods and the KD-enhanced variants (i.e., KD + Fine-Tune/LWC-KD-PIW).
Overall, methods that leverage KD from the teacher systems (i.e., KD + Fine-Tune/LWC-KD-PIW, \proposed) achieve higher performance than the remaining methods training the student by itself.
This shows the importance of leveraging extensive teacher knowledge.
Moreover, we observe that the effectiveness of the CL methods is rather limited for the student models, particularly for the latter blocks (i.e., D4 and D5).
The student has a highly limited capacity and learning capability, making it more challenging to update effectively.
This aspect has not been explicitly considered in previous CL methods. 
\proposed introduces two new strategies, entity embedding initialization and proxy-guided replay learning, which effectively support the student's learning process.

It is also noted that \proposed even generally outperforms Full-Batch in terms of LA.
Full-Batch is certainly the strongest baseline for RA (stability), as it learns from all observed historical data. 
However, it is not always the strongest method for LA (plasticity), as it treats all data equally without putting emphasis on the latest interactions. 
\proposed effectively learns the latest interactions while preserving previous knowledge, which leads to enhanced performance~in~both~LA~and~RA.



\smallsection{Collaborative evolution.}
Figure \ref{fig:gap_images} presents the absolute H-mean gain of \proposed over the best CL competitor.
We report the results of Recall@20 with the GNN-based student.
In \proposed, the teacher and the student collaboratively improve each other by mutual knowledge exchange (\cref{sec:s1}, \cref{sec:s3}).
These improvements accumulate over time, progressively refining both the teacher and student models. 
As a result, the performance gap between \proposed and the CL competitor generally widens over time, which ascertains the collaborative evolution~in~\proposed.


\subsubsection{\textbf{Accuracy-efficiency comparison.}}
Table \ref{tab:abl_efficiency} compares the accuracy and efficiency aspects of teacher and student, trained by the best CL method (i.e., LWC-KD-PIW) and \proposed, respectively.
We increase the student size until it achieves \textit{comparable} performance to the teacher.
Then, we report the average H-mean (Recall@20) for all blocks, the number of parameters, and the time required for training and inference.
We utilize PyTorch with CUDA from RTX A6000 and AMD EPYC 7313 CPU.
Compared to the massive teacher system which has significant computational costs for training and consolidating the multiple large-scale models, the compact student model trained with \proposed significantly reduces the huge computational burdens while maintaining the high performance of the teacher system.

Lastly, we highlight that the KD process takes negligible time compared to the teacher update. 
Specifically, on the Yelp dataset, KD takes about 8 minutes, whereas the teacher update takes about 107 minutes. 
Considering the student update takes about 5 minutes, the KD process, followed by student updates, is more efficient than repetitive updates of the large-scale teacher, making it more suitable for real-time applications.


\begin{table}[t]
\centering
\caption{Accuracy and efficiency comparison. Train time refers to the time required to train the model until convergence. Inference time indicates the average wall time for generating recommendations for all users.}
\label{tab:abl_efficiency}
\renewcommand{\tabcolsep}{1.1mm}
\resizebox{\linewidth}{!}{
\begin{tabular}{c|cl|cccc}
\hline
\textbf{Dataset} &  & \textbf{Method} & \textbf{H-mean} & \multicolumn{1}{c}{\textbf{\#Params.}} & \textbf{Train time} & \textbf{Inference time} \\
 \hline \hline
\multirow{2}{*}{Gowalla} & Teacher & LWC-KD-PIW & 0.1332 & 33.47M & 14h 53m 2s & 301.22s \\
& Student & \proposed & 0.1303 & 2.71M & 52m 54s & 49.14s \\
 \hline
\multirow{2}{*}{Yelp}& Teacher & LWC-KD-PIW & 0.0813 & 18.24M & 1h 47m 47s & 105.20s \\
& Student & \proposed & 0.0861 & 0.99M & 5m 34s & 6.56s \\
 \hline
\end{tabular}}
\end{table}



\subsection{Study of \proposed}
We provide comprehensive analyses for an in-depth understanding of \proposed.
We report the results with the GNN-based student on~Yelp.

\subsubsection{\textbf{Ablation study.}}
We present the ablation study of CCD to demonstrate the effectiveness of each proposed component.

\smallsection{Teacher side.}
Table \ref{tab:abl_teacher} presents the ablation results on the teacher.
From \cref{sec:s3}, we exclude three proposed components:
(1) `w/o student-side knowledge' excludes $\mathcal{L}_{\text{S} \rightarrow \text{T}}$, (2) `w/o proxies (student only)' only uses the updated student excluding the proxies in $\mathcal{L}_{\text{S} \rightarrow \text{T}}$, and (3) `w/o annealing' excludes the loss annealing.
First, excluding the student-side knowledge largely degrades the effectiveness of \proposed, and the best performance is achieved by leveraging both the updated student and the proxies accumulating the historical knowledge.
Also, the simple annealing that gradually reduces the impacts of the student-side knowledge effectively improves the~teacher.


\begin{table}[t]
\centering
\caption{Results of various ablations on teacher system.}
\renewcommand{\tabcolsep}{1.1mm}
\label{tab:abl_teacher}
\resizebox{\linewidth}{!}{
\begin{tabular}{l|ccc|ccc}
\hline
 & \multicolumn{3}{c|}{\textbf{After $D_4$}}          & \multicolumn{3}{c}{\textbf{After $D_5$}}            \\
\multirow{-2}{*}{{\textbf{Method}}} & \textbf{LA}     & \textbf{RA}    & \textbf{H-mean} & \textbf{LA}     & \textbf{RA}     & \textbf{H-mean} \\ \hline\hline
\textbf{\proposed}        & \textbf{0.1209} & \textbf{0.0950} & \textbf{0.1064} & \textbf{0.1198} & \textbf{0.0914} & \textbf{0.1037} \\ \hline
w/o student-side knowledge                                                    & 0.1083          & 0.0491         & 0.0676          & 0.1075          & 0.0499          & 0.0682          \\
w/o proxies (student only)                                                & 0.0933          & 0.0683         & 0.0789          & 0.0956          & 0.0666          & 0.0785          \\
w/o annealing  &0.1021	&0.0645	&0.0791  &0.1034 &0.0649	&0.0797          \\\hline
\end{tabular}
}\vspace{-0.4cm}
\end{table}

\smallsection{Student side.}
Table \ref{tab:abl_student} presents the ablation results on the student.
First, we examine the effects of proxy-guided replay learning (\cref{subsub:s_proxy}).
`w/o proxy learning' excludes $\mathcal{L}_{RE}$, and `w/o S-/P-proxy' exclude the stability proxy and the plasticity proxy, respectively.
We observe that both proxies are indeed beneficial for the student, as they provide complementary views of the previous knowledge, as shown in Figure \ref{fig:proxy}.
Second, we investigate the effect of $w_{SP}$ and $w_{PP}$, used to update the proxies in Eq.\ref{eq:proxy}.
We assess the extreme case for each proxy.
When $w_{SP}=0$, the stability proxy does not accept any recent knowledge (i.e., extreme stability).
Conversely, when $w_{PP}=1$, the plasticity proxy is entirely replaced by the recent knowledge (i.e., extreme plasticity).
The best performance is achieved when both proxies are updated with an appropriate balance.
However, as long as the proxies accumulate past knowledge, the choice of update weights has little impact on the final performance. We set $w_{SP} = 0.1$ and $w_{PP} = 0.9$, respectively.


\begin{table}[t]
\centering
\caption{Results of various ablations on student model.}
\renewcommand{\tabcolsep}{0.8mm}
\label{tab:abl_student}
\resizebox{\linewidth}{!}{
\begin{tabular}{cl|ccc|ccc}
\hline
\multicolumn{2}{c|}{\multirow{2}{*}{\textbf{Method}}}  & \multicolumn{3}{c|}{\textbf{After $D_4$}}    & \multicolumn{3}{c}{\textbf{After $D_5$}}       \\
\multicolumn{2}{c|}{}       & \textbf{LA}     & \textbf{RA}     & \textbf{H-mean} & \textbf{LA}     & \textbf{RA}     & \textbf{H-mean} \\ \hline\hline
\multicolumn{2}{c|}{\textbf{\proposed}}   & \textbf{0.1090} & \textbf{0.0913} & \textbf{0.0994} & \textbf{0.1100} & \textbf{0.0878} & \textbf{0.0977} \\ \hline
\multicolumn{1}{c|}{\multirow{3}{*}{\textbf{\begin{tabular}[c]{@{}c@{}}Proxy-guided \\ learning \end{tabular}}}}   & w/o proxy learning & 0.0972  & 0.0656  & 0.0783  & 0.0965  & 0.0596  & 0.0737  \\
\multicolumn{1}{c|}{}  & w/o S-proxy        & 0.0959  & 0.0713  & 0.0818  & 0.0965  & 0.0727  & 0.0829  \\
\multicolumn{1}{c|}{}  & w/o P-proxy        & 0.0981  & 0.0694  & 0.0813  & 0.0981  & 0.0713  & 0.0826  \\ \hline
\multicolumn{1}{c|}{\multirow{2}{*}{\textbf{\begin{tabular}[c]{@{}c@{}}Proxy \\update \end{tabular}}}} & extreme stability  & 0.1035  & 0.0904  & 0.0965  & 0.1052  & 0.0869  & 0.0952  \\
\multicolumn{1}{c|}{}  & extreme plasticity  & 0.1062  & 0.0910  & 0.0980   & 0.1075  & 0.0871  & 0.0962  \\
\hline
\end{tabular}}
\vspace{0.2cm}
\end{table}


\subsubsection{\textbf{Impacts of the replay size.}}
\begin{figure}[t]
  \includegraphics[width=0.4\columnwidth]{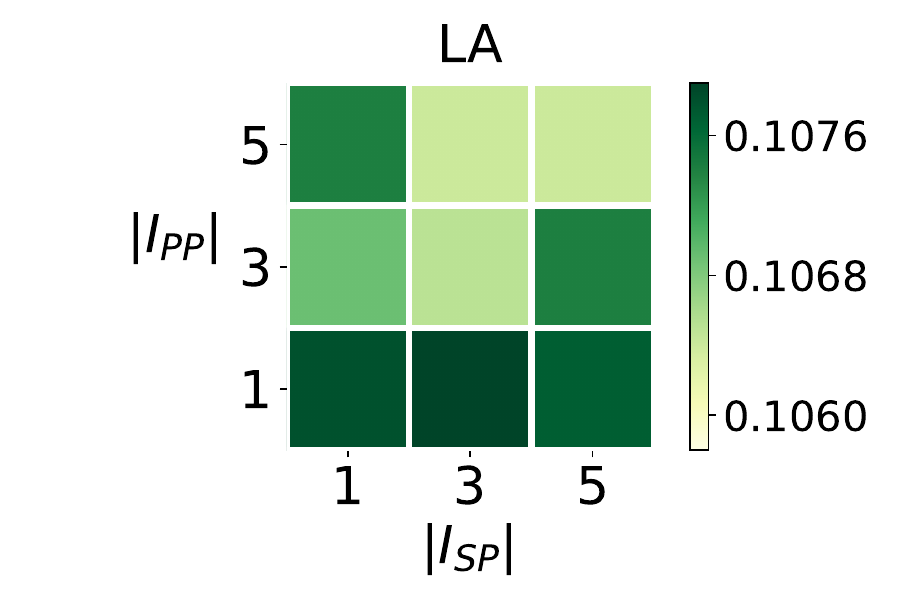}
  \includegraphics[width=0.4\columnwidth]{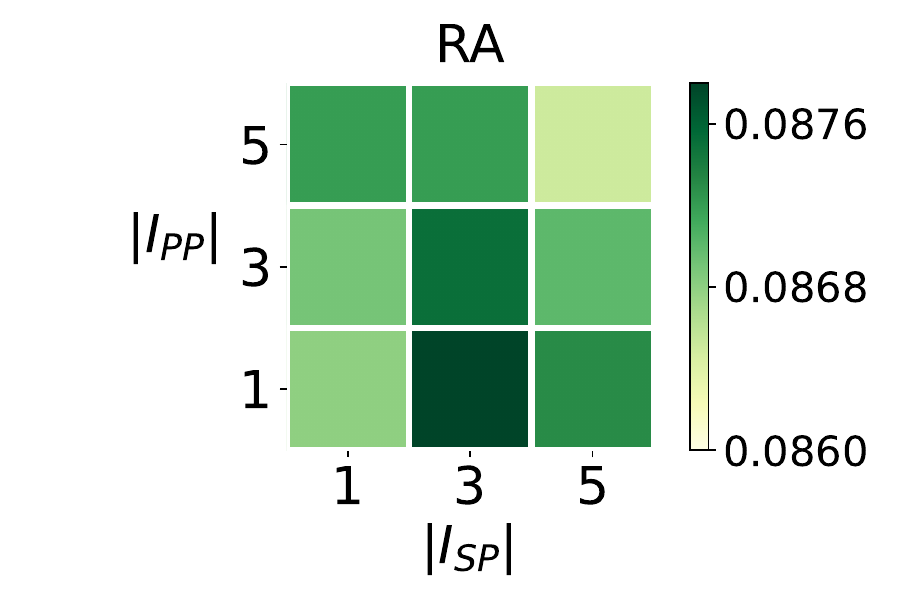}
  \caption{Performance with varying replay sizes.} 
  \label{fig:hyperparam}
  \vspace{-0.3cm}
\end{figure}

Figure \ref{fig:hyperparam} presents the performance (Recall@20) of the student model trained with varying replay sizes in Eq.\ref{eq:pd}.
The best performance is achieved with a small sampling size ($\leq 3$), suggesting that our proxy-guided replay learning does not notably increase the training complexity.

\subsubsection{\textbf{Analysis on dormant/new users.}}
We further analyze the recommendation quality of the student for specific user groups having distinct characteristics.
In Table \ref{tab:longtime}, we present the recommendation performance for \textit{dormant users} who haven't used the system for an extended period. 
We select users who were active in $D_1$, remained dormant from $D_2$ to $D_4$, and became active again in $D_5$. 
We evaluate performance for the users on the test set of $D_5$, after updating the student up to $D_4$.
We compare the baselines that show high effectiveness in previous experiments.
We observe that \proposed generates more accurate recommendations for the dormant users, indicating that \proposed effectively preserves previous~knowledge.

In Table \ref{tab:newuser}, we present the average recommendation performance for \textit{new users} for each data block.
\proposed proposes to use the prominent-entity information to facilitate the understanding of new entities (\cref{subsub:s_CF}).
To assess its effectiveness, we report the results when replacing the proposed technique with conventional random initialization and the widely used 1-hop initialization \cite{inmo, inductiveMC_CIKM21}.
The proposed technique effectively enhances the student's ability to adapt to new entities.
Moreover, as the teacher subsequently leverages the student-side knowledge, it leads to further improvement in teacher performance.
These results collectively show that \proposed achieves a good balance between stability and~plasticity.

\begin{table}[t]
\caption{Recommendation performance on dormant users.}
\label{tab:longtime}
\renewcommand{\arraystretch}{0.75}
\resizebox{0.65\linewidth}{!}{
\begin{tabular}{c|cc}\hline
\textbf{Method} & \textbf{Recall@20} & \textbf{NDCG@20} \\\hline\hline
LWC-KD-PIW & 0.0081 & 0.0024 \\
KD + Fine-Tune & 0.0645 & 0.0286 \\
\proposed & \textbf{0.0806} &\textbf{0.0311} \\ \hline
\end{tabular}}
\vspace{-0.3cm}
\end{table}

\begin{table}[t]
\caption{Effects of new entity initialization on new users.}
\label{tab:newuser}
\renewcommand{\arraystretch}{0.75}
\resizebox{0.99\linewidth}{!}{
\begin{tabular}{l | cc|cc} \hline
 & \multicolumn{2}{c|}{\textbf{Student}} & \multicolumn{2}{c}{\textbf{Teacher}} \\
 &\textbf{ Recall@20} & \textbf{NDCG@20} & \textbf{Recall@20} & \textbf{NDCG@20} \\ \hline\hline
\textbf{\proposed} & \textbf{0.1313} & \textbf{0.0559} & \textbf{0.1500} & \textbf{0.0605}\\ \hline
w/ random init. & 0.1270 & 0.0546 & 0.1387 & 0.0573 \\
w/ one-hop init. & 0.1116 & 0.0436 & 0.1037 & 0.0420 \\\hline
\end{tabular}}
\end{table}

\section{Conclusion}
\label{sec:conclusion}
We propose \proposed framework for effective and efficient deployment through a compact student model having the high performance of the massive teacher system, while aptly adapting to continuously incoming data.
Unlike the existing KD studies have focused on one-time distillation in static environments, \proposed updates both the teacher and the student continually and collaboratively along the data stream.
\proposed facilitates the student's effective adaptation to new data, while also enabling the teacher to fully leverage accumulated knowledge.
Given its great compatibility with existing models, we expect that our \proposed framework can provide a better accuracy-efficiency trade-off for practical recommender systems.

\section*{Acknowledgement}
This work was supported by the IITP grant funded by the MSIT (South Korea, No.2018-0-00584, RS-2019-II191906), the NRF grant funded by the MSIT (South Korea, No.RS-2023-00217286, No.2020R1\\A2B5B03097210), the TIP funded by the MOTIE (South Korea, No.200\\14926), and the DIP grant funded by the MSIT and Daegu Metropolitan City (South Korea, No. DBSD1-07).


\clearpage

\bibliographystyle{ACM-Reference-Format}
\balance
\bibliography{acmart}


\begin{thebibliography}{50}


\ifx \showCODEN    \undefined \def \showCODEN     #1{\unskip}     \fi
\ifx \showDOI      \undefined \def \showDOI       #1{#1}\fi
\ifx \showISBNx    \undefined \def \showISBNx     #1{\unskip}     \fi
\ifx \showISBNxiii \undefined \def \showISBNxiii  #1{\unskip}     \fi
\ifx \showISSN     \undefined \def \showISSN      #1{\unskip}     \fi
\ifx \showLCCN     \undefined \def \showLCCN      #1{\unskip}     \fi
\ifx \shownote     \undefined \def \shownote      #1{#1}          \fi
\ifx \showarticletitle \undefined \def \showarticletitle #1{#1}   \fi
\ifx \showURL      \undefined \def \showURL       {\relax}        \fi
\providecommand\bibfield[2]{#2}
\providecommand\bibinfo[2]{#2}
\providecommand\natexlab[1]{#1}
\providecommand\showeprint[2][]{arXiv:#2}

\bibitem[Ahrabian et~al\mbox{.}(2021)]%
        {CL_RS_replay}
\bibfield{author}{\bibinfo{person}{Kian Ahrabian}, \bibinfo{person}{Yishi Xu}, \bibinfo{person}{Yingxue Zhang}, \bibinfo{person}{Jiapeng Wu}, \bibinfo{person}{Yuening Wang}, {and} \bibinfo{person}{Mark Coates}.} \bibinfo{year}{2021}\natexlab{}.
\newblock \showarticletitle{Structure aware experience replay for incremental learning in graph-based recommender systems}. In \bibinfo{booktitle}{\emph{CIKM}}. \bibinfo{pages}{2832--2836}.
\newblock


\bibitem[Bao et~al\mbox{.}(2023)]%
        {bao2023tallrec}
\bibfield{author}{\bibinfo{person}{Keqin Bao}, \bibinfo{person}{Jizhi Zhang}, \bibinfo{person}{Yang Zhang}, \bibinfo{person}{Wenjie Wang}, \bibinfo{person}{Fuli Feng}, {and} \bibinfo{person}{Xiangnan He}.} \bibinfo{year}{2023}\natexlab{}.
\newblock \showarticletitle{Tallrec: An effective and efficient tuning framework to align large language model with recommendation}. In \bibinfo{booktitle}{\emph{Proceedings of the 17th ACM Conference on Recommender Systems}}. \bibinfo{pages}{1007--1014}.
\newblock


\bibitem[Cai et~al\mbox{.}(2022)]%
        {cai2022reloop}
\bibfield{author}{\bibinfo{person}{Guohao Cai}, \bibinfo{person}{Jieming Zhu}, \bibinfo{person}{Quanyu Dai}, \bibinfo{person}{Zhenhua Dong}, \bibinfo{person}{Xiuqiang He}, \bibinfo{person}{Ruiming Tang}, {and} \bibinfo{person}{Rui Zhang}.} \bibinfo{year}{2022}\natexlab{}.
\newblock \showarticletitle{ReLoop: A Self-Correction Continual Learning Loop for Recommender Systems}. In \bibinfo{booktitle}{\emph{SIGIR}}. \bibinfo{pages}{2692--2697}.
\newblock


\bibitem[Chen et~al\mbox{.}(2022)]%
        {chen2022learning}
\bibfield{author}{\bibinfo{person}{Yankai Chen}, \bibinfo{person}{Huifeng Guo}, \bibinfo{person}{Yingxue Zhang}, \bibinfo{person}{Chen Ma}, \bibinfo{person}{Ruiming Tang}, \bibinfo{person}{Jingjie Li}, {and} \bibinfo{person}{Irwin King}.} \bibinfo{year}{2022}\natexlab{}.
\newblock \showarticletitle{Learning Binarized Graph Representations with Multi-faceted Quantization Reinforcement for Top-K Recommendation}. In \bibinfo{booktitle}{\emph{KDD}}.
\newblock


\bibitem[Do and Lauw(2023)]%
        {do2023continual}
\bibfield{author}{\bibinfo{person}{Jaime~Hieu Do} {and} \bibinfo{person}{Hady~W Lauw}.} \bibinfo{year}{2023}\natexlab{}.
\newblock \showarticletitle{Continual Collaborative Filtering Through Gradient Alignment}. In \bibinfo{booktitle}{\emph{RecSys}}. \bibinfo{pages}{1133--1138}.
\newblock


\bibitem[Du et~al\mbox{.}(2021)]%
        {du2021alternate}
\bibfield{author}{\bibinfo{person}{Xiaocong Du}, \bibinfo{person}{Bhargav Bhushanam}, \bibinfo{person}{Jiecao Yu}, \bibinfo{person}{Dhruv Choudhary}, \bibinfo{person}{Tianxiang Gao}, \bibinfo{person}{Sherman Wong}, \bibinfo{person}{Louis Feng}, \bibinfo{person}{Jongsoo Park}, \bibinfo{person}{Yu Cao}, {and} \bibinfo{person}{Arun Kejariwal}.} \bibinfo{year}{2021}\natexlab{}.
\newblock \showarticletitle{Alternate model growth and pruning for efficient training of recommendation systems}. In \bibinfo{booktitle}{\emph{20th IEEE International Conference on Machine Learning and Applications (ICMLA)}}. IEEE, \bibinfo{pages}{1421--1428}.
\newblock


\bibitem[Geng et~al\mbox{.}(2022)]%
        {geng2022recommendation}
\bibfield{author}{\bibinfo{person}{Shijie Geng}, \bibinfo{person}{Shuchang Liu}, \bibinfo{person}{Zuohui Fu}, \bibinfo{person}{Yingqiang Ge}, {and} \bibinfo{person}{Yongfeng Zhang}.} \bibinfo{year}{2022}\natexlab{}.
\newblock \showarticletitle{Recommendation as language processing (rlp): A unified pretrain, personalized prompt \& predict paradigm (p5)}. In \bibinfo{booktitle}{\emph{RecSys}}. \bibinfo{pages}{299--315}.
\newblock


\bibitem[He et~al\mbox{.}(2020)]%
        {he2020lightgcn}
\bibfield{author}{\bibinfo{person}{Xiangnan He}, \bibinfo{person}{Kuan Deng}, \bibinfo{person}{Xiang Wang}, \bibinfo{person}{Yan Li}, \bibinfo{person}{Yongdong Zhang}, {and} \bibinfo{person}{Meng Wang}.} \bibinfo{year}{2020}\natexlab{}.
\newblock \showarticletitle{LightGCN: Simplifying and Powering Graph Convolution Network for Recommendation}. In \bibinfo{booktitle}{\emph{SIGIR}}.
\newblock


\bibitem[He et~al\mbox{.}(2017)]%
        {NeuMF}
\bibfield{author}{\bibinfo{person}{Xiangnan He}, \bibinfo{person}{Lizi Liao}, \bibinfo{person}{Hanwang Zhang}, \bibinfo{person}{Liqiang Nie}, \bibinfo{person}{Xia Hu}, {and} \bibinfo{person}{Tat-Seng Chua}.} \bibinfo{year}{2017}\natexlab{}.
\newblock \showarticletitle{Neural collaborative filtering}. In \bibinfo{booktitle}{\emph{WWW}}.
\newblock


\bibitem[Hinton et~al\mbox{.}(2015)]%
        {KD}
\bibfield{author}{\bibinfo{person}{Geoffrey Hinton}, \bibinfo{person}{Oriol Vinyals}, {and} \bibinfo{person}{Jeffrey Dean}.} \bibinfo{year}{2015}\natexlab{}.
\newblock \showarticletitle{Distilling the knowledge in a neural network}. In \bibinfo{booktitle}{\emph{NeurIPS}}.
\newblock


\bibitem[Hsieh et~al\mbox{.}(2017)]%
        {CML}
\bibfield{author}{\bibinfo{person}{Cheng-Kang Hsieh}, \bibinfo{person}{Longqi Yang}, \bibinfo{person}{Yin Cui}, \bibinfo{person}{Tsung-Yi Lin}, \bibinfo{person}{Serge Belongie}, {and} \bibinfo{person}{Deborah Estrin}.} \bibinfo{year}{2017}\natexlab{}.
\newblock \showarticletitle{Collaborative metric learning}. In \bibinfo{booktitle}{\emph{WWW}}.
\newblock


\bibitem[Kang et~al\mbox{.}(2020)]%
        {DERRD}
\bibfield{author}{\bibinfo{person}{SeongKu Kang}, \bibinfo{person}{Junyoung Hwang}, \bibinfo{person}{Wonbin Kweon}, {and} \bibinfo{person}{Hwanjo Yu}.} \bibinfo{year}{2020}\natexlab{}.
\newblock \showarticletitle{DE-RRD: A Knowledge Distillation Framework for Recommender System}. In \bibinfo{booktitle}{\emph{CIKM}}.
\newblock


\bibitem[Kang et~al\mbox{.}(2021a)]%
        {IRRRD}
\bibfield{author}{\bibinfo{person}{SeongKu Kang}, \bibinfo{person}{Junyoung Hwang}, \bibinfo{person}{Wonbin Kweon}, {and} \bibinfo{person}{Hwanjo Yu}.} \bibinfo{year}{2021}\natexlab{a}.
\newblock \showarticletitle{Item-side ranking regularized distillation for recommender system}.
\newblock \bibinfo{journal}{\emph{Information Sciences}}  \bibinfo{volume}{580} (\bibinfo{year}{2021}), \bibinfo{pages}{15--34}.
\newblock
\showISSN{0020-0255}
\urldef\tempurl%
\url{https://doi.org/10.1016/j.ins.2021.08.060}
\showDOI{\tempurl}


\bibitem[Kang et~al\mbox{.}(2021b)]%
        {TD}
\bibfield{author}{\bibinfo{person}{SeongKu Kang}, \bibinfo{person}{Junyoung Hwang}, \bibinfo{person}{Wonbin Kweon}, {and} \bibinfo{person}{Hwanjo Yu}.} \bibinfo{year}{2021}\natexlab{b}.
\newblock \showarticletitle{Topology Distillation for Recommender System}. In \bibinfo{booktitle}{\emph{KDD}}.
\newblock


\bibitem[Kang et~al\mbox{.}(2019)]%
        {SSCDR}
\bibfield{author}{\bibinfo{person}{SeongKu Kang}, \bibinfo{person}{Junyoung Hwang}, \bibinfo{person}{Dongha Lee}, {and} \bibinfo{person}{Hwanjo Yu}.} \bibinfo{year}{2019}\natexlab{}.
\newblock \showarticletitle{Semi-supervised learning for cross-domain recommendation to cold-start users}. In \bibinfo{booktitle}{\emph{CIKM}}.
\newblock


\bibitem[Kang et~al\mbox{.}(2023)]%
        {hetcomp}
\bibfield{author}{\bibinfo{person}{SeongKu Kang}, \bibinfo{person}{Wonbin Kweon}, \bibinfo{person}{Dongha Lee}, \bibinfo{person}{Jianxun Lian}, \bibinfo{person}{Xing Xie}, {and} \bibinfo{person}{Hwanjo Yu}.} \bibinfo{year}{2023}\natexlab{}.
\newblock \showarticletitle{Distillation from Heterogeneous Models for Top-K Recommendation}. In \bibinfo{booktitle}{\emph{WWW}}. \bibinfo{pages}{801--811}.
\newblock


\bibitem[Kang et~al\mbox{.}(2024)]%
        {Kang_unbiased}
\bibfield{author}{\bibinfo{person}{SeongKu Kang}, \bibinfo{person}{Wonbin Kweon}, \bibinfo{person}{Dongha Lee}, \bibinfo{person}{Jianxun Lian}, \bibinfo{person}{Xing Xie}, {and} \bibinfo{person}{Hwanjo Yu}.} \bibinfo{year}{2024}\natexlab{}.
\newblock \showarticletitle{Unbiased, Effective, and Efficient Distillation from Heterogeneous Models for Recommender Systems}.
\newblock \bibinfo{journal}{\emph{ACM Trans. Recomm. Syst.}} (\bibinfo{date}{feb} \bibinfo{year}{2024}).
\newblock
\urldef\tempurl%
\url{https://doi.org/10.1145/3649443}
\showDOI{\tempurl}


\bibitem[Kang et~al\mbox{.}(2022b)]%
        {concf}
\bibfield{author}{\bibinfo{person}{SeongKu Kang}, \bibinfo{person}{Dongha Lee}, \bibinfo{person}{Wonbin Kweon}, \bibinfo{person}{Junyoung Hwang}, {and} \bibinfo{person}{Hwanjo Yu}.} \bibinfo{year}{2022}\natexlab{b}.
\newblock \showarticletitle{Consensus Learning from Heterogeneous Objectives for One-Class Collaborative Filtering}. In \bibinfo{booktitle}{\emph{WWW}}.
\newblock


\bibitem[Kang et~al\mbox{.}(2022a)]%
        {PHR}
\bibfield{author}{\bibinfo{person}{SeongKu Kang}, \bibinfo{person}{Dongha Lee}, \bibinfo{person}{Wonbin Kweon}, {and} \bibinfo{person}{Hwanjo Yu}.} \bibinfo{year}{2022}\natexlab{a}.
\newblock \showarticletitle{Personalized Knowledge Distillation for Recommender System}.
\newblock \bibinfo{journal}{\emph{Knowledge-Based Systems}}  \bibinfo{volume}{239} (\bibinfo{year}{2022}), \bibinfo{pages}{107958}.
\newblock
\showISSN{0950-7051}
\urldef\tempurl%
\url{https://doi.org/10.1016/j.knosys.2021.107958}
\showDOI{\tempurl}


\bibitem[Kirkpatrick et~al\mbox{.}(2017)]%
        {kirkpatrick2017overcoming}
\bibfield{author}{\bibinfo{person}{James Kirkpatrick}, \bibinfo{person}{Razvan Pascanu}, \bibinfo{person}{Neil Rabinowitz}, \bibinfo{person}{Joel Veness}, \bibinfo{person}{Guillaume Desjardins}, \bibinfo{person}{Andrei~A Rusu}, \bibinfo{person}{Kieran Milan}, \bibinfo{person}{John Quan}, \bibinfo{person}{Tiago Ramalho}, \bibinfo{person}{Agnieszka Grabska-Barwinska}, {et~al\mbox{.}}} \bibinfo{year}{2017}\natexlab{}.
\newblock \showarticletitle{Overcoming catastrophic forgetting in neural networks}.
\newblock \bibinfo{journal}{\emph{Proceedings of the national academy of sciences}} \bibinfo{volume}{114}, \bibinfo{number}{13} (\bibinfo{year}{2017}), \bibinfo{pages}{3521--3526}.
\newblock


\bibitem[Kweon et~al\mbox{.}(2021)]%
        {BD}
\bibfield{author}{\bibinfo{person}{Wonbin Kweon}, \bibinfo{person}{SeongKu Kang}, {and} \bibinfo{person}{Hwanjo Yu}.} \bibinfo{year}{2021}\natexlab{}.
\newblock \showarticletitle{Bidirectional Distillation for Top-K Recommender System}. In \bibinfo{booktitle}{\emph{WWW}}.
\newblock


\bibitem[Lee et~al\mbox{.}(2021)]%
        {BUIR}
\bibfield{author}{\bibinfo{person}{Dongha Lee}, \bibinfo{person}{SeongKu Kang}, \bibinfo{person}{Hyunjun Ju}, \bibinfo{person}{Chanyoung Park}, {and} \bibinfo{person}{Hwanjo Yu}.} \bibinfo{year}{2021}\natexlab{}.
\newblock \showarticletitle{Bootstrapping User and Item Representations for One-Class Collaborative Filtering}. In \bibinfo{booktitle}{\emph{SIGIR}}.
\newblock


\bibitem[Lee et~al\mbox{.}(2023)]%
        {lee2023mvfs}
\bibfield{author}{\bibinfo{person}{Youngjune Lee}, \bibinfo{person}{Yeongjong Jeong}, \bibinfo{person}{Keunchan Park}, {and} \bibinfo{person}{SeongKu Kang}.} \bibinfo{year}{2023}\natexlab{}.
\newblock \showarticletitle{MvFS: Multi-view Feature Selection for Recommender System}. In \bibinfo{booktitle}{\emph{CIKM}}. \bibinfo{pages}{4048--4052}.
\newblock


\bibitem[Lee and Kim(2021)]%
        {DCD}
\bibfield{author}{\bibinfo{person}{Youngjune Lee} {and} \bibinfo{person}{Kee-Eung Kim}.} \bibinfo{year}{2021}\natexlab{}.
\newblock \showarticletitle{Dual Correction Strategy for Ranking Distillation in Top-N Recommender System}. In \bibinfo{booktitle}{\emph{CIKM}}.
\newblock


\bibitem[Liang et~al\mbox{.}(2018)]%
        {VAE}
\bibfield{author}{\bibinfo{person}{Dawen Liang}, \bibinfo{person}{Rahul~G. Krishnan}, \bibinfo{person}{Matthew~D. Hoffman}, {and} \bibinfo{person}{Tony Jebara}.} \bibinfo{year}{2018}\natexlab{}.
\newblock \showarticletitle{Variational Autoencoders for Collaborative Filtering}. In \bibinfo{booktitle}{\emph{WWW}}.
\newblock


\bibitem[Lin et~al\mbox{.}(2022)]%
        {lin2022towards}
\bibfield{author}{\bibinfo{person}{Guoliang Lin}, \bibinfo{person}{Hanlu Chu}, {and} \bibinfo{person}{Hanjiang Lai}.} \bibinfo{year}{2022}\natexlab{}.
\newblock \showarticletitle{Towards better plasticity-stability trade-off in incremental learning: A simple linear connector}. In \bibinfo{booktitle}{\emph{CVPR}}. \bibinfo{pages}{89--98}.
\newblock


\bibitem[Marden(1996)]%
        {marden1996analyzing}
\bibfield{author}{\bibinfo{person}{John~I Marden}.} \bibinfo{year}{1996}\natexlab{}.
\newblock \bibinfo{booktitle}{\emph{Analyzing and modeling rank data}}.
\newblock \bibinfo{publisher}{CRC Press}.
\newblock


\bibitem[McClelland et~al\mbox{.}(1995)]%
        {CLS}
\bibfield{author}{\bibinfo{person}{James~L McClelland}, \bibinfo{person}{Bruce~L McNaughton}, {and} \bibinfo{person}{Randall~C O'Reilly}.} \bibinfo{year}{1995}\natexlab{}.
\newblock \showarticletitle{Why there are complementary learning systems in the hippocampus and neocortex: insights from the successes and failures of connectionist models of learning and memory.}
\newblock \bibinfo{journal}{\emph{Psychological review}} \bibinfo{volume}{102}, \bibinfo{number}{3} (\bibinfo{year}{1995}), \bibinfo{pages}{419}.
\newblock


\bibitem[Mi et~al\mbox{.}(2020)]%
        {mi2020ader}
\bibfield{author}{\bibinfo{person}{Fei Mi}, \bibinfo{person}{Xiaoyu Lin}, {and} \bibinfo{person}{Boi Faltings}.} \bibinfo{year}{2020}\natexlab{}.
\newblock \showarticletitle{Ader: Adaptively distilled exemplar replay towards continual learning for session-based recommendation}. In \bibinfo{booktitle}{\emph{RecSys}}. \bibinfo{pages}{408--413}.
\newblock


\bibitem[Pham et~al\mbox{.}(2021)]%
        {dualnet}
\bibfield{author}{\bibinfo{person}{Quang Pham}, \bibinfo{person}{Chenghao Liu}, {and} \bibinfo{person}{Steven Hoi}.} \bibinfo{year}{2021}\natexlab{}.
\newblock \showarticletitle{Dualnet: Continual learning, fast and slow}. In \bibinfo{booktitle}{\emph{NeurIPS}}. \bibinfo{pages}{16131--16144}.
\newblock


\bibitem[Reddi et~al\mbox{.}(2021)]%
        {reddi2021rankdistil}
\bibfield{author}{\bibinfo{person}{Sashank Reddi}, \bibinfo{person}{Rama~Kumar Pasumarthi}, \bibinfo{person}{Aditya Menon}, \bibinfo{person}{Ankit~Singh Rawat}, \bibinfo{person}{Felix Yu}, \bibinfo{person}{Seungyeon Kim}, \bibinfo{person}{Andreas Veit}, {and} \bibinfo{person}{Sanjiv Kumar}.} \bibinfo{year}{2021}\natexlab{}.
\newblock \showarticletitle{Rankdistil: Knowledge distillation for ranking}. In \bibinfo{booktitle}{\emph{AISTATS}}. PMLR.
\newblock


\bibitem[Rendle et~al\mbox{.}(2009)]%
        {BPR}
\bibfield{author}{\bibinfo{person}{Steffen Rendle}, \bibinfo{person}{Christoph Freudenthaler}, \bibinfo{person}{Zeno Gantner}, {and} \bibinfo{person}{Lars Schmidt-Thieme}.} \bibinfo{year}{2009}\natexlab{}.
\newblock \showarticletitle{BPR: Bayesian personalized ranking from implicit feedback}. In \bibinfo{booktitle}{\emph{UAI}}.
\newblock


\bibitem[Romero et~al\mbox{.}(2015)]%
        {FitNet}
\bibfield{author}{\bibinfo{person}{Adriana Romero}, \bibinfo{person}{Nicolas Ballas}, \bibinfo{person}{Samira~Ebrahimi Kahou}, \bibinfo{person}{Antoine Chassang}, \bibinfo{person}{Carlo Gatta}, {and} \bibinfo{person}{Yoshua Bengio}.} \bibinfo{year}{2015}\natexlab{}.
\newblock \showarticletitle{Fitnets: Hints for thin deep nets}. In \bibinfo{booktitle}{\emph{ICLR}}.
\newblock


\bibitem[Shen et~al\mbox{.}(2021)]%
        {inductiveMC_CIKM21}
\bibfield{author}{\bibinfo{person}{Wei Shen}, \bibinfo{person}{Chuheng Zhang}, \bibinfo{person}{Yun Tian}, \bibinfo{person}{Liang Zeng}, \bibinfo{person}{Xiaonan He}, \bibinfo{person}{Wanchun Dou}, {and} \bibinfo{person}{Xiaolong Xu}.} \bibinfo{year}{2021}\natexlab{}.
\newblock \showarticletitle{Inductive Matrix Completion Using Graph Autoencoder}. In \bibinfo{booktitle}{\emph{CIKM}}. \bibinfo{pages}{1609--1618}.
\newblock


\bibitem[Shin et~al\mbox{.}(2017)]%
        {shin2017continual}
\bibfield{author}{\bibinfo{person}{Hanul Shin}, \bibinfo{person}{Jung~Kwon Lee}, \bibinfo{person}{Jaehong Kim}, {and} \bibinfo{person}{Jiwon Kim}.} \bibinfo{year}{2017}\natexlab{}.
\newblock \showarticletitle{Continual learning with deep generative replay}. In \bibinfo{booktitle}{\emph{NeurIPS}}.
\newblock


\bibitem[Tang and Wang(2018)]%
        {RD}
\bibfield{author}{\bibinfo{person}{Jiaxi Tang} {and} \bibinfo{person}{Ke Wang}.} \bibinfo{year}{2018}\natexlab{}.
\newblock \showarticletitle{Ranking distillation: Learning compact ranking models with high performance for recommender system}. In \bibinfo{booktitle}{\emph{KDD}}.
\newblock


\bibitem[Tarvainen and Valpola(2017)]%
        {mean_teacher}
\bibfield{author}{\bibinfo{person}{Antti Tarvainen} {and} \bibinfo{person}{Harri Valpola}.} \bibinfo{year}{2017}\natexlab{}.
\newblock \showarticletitle{Mean teachers are better role models: Weight-averaged consistency targets improve semi-supervised deep learning results}. In \bibinfo{booktitle}{\emph{NeurIPS}}.
\newblock


\bibitem[Wang et~al\mbox{.}(2020)]%
        {wang2020m2grl}
\bibfield{author}{\bibinfo{person}{Menghan Wang}, \bibinfo{person}{Yujie Lin}, \bibinfo{person}{Guli Lin}, \bibinfo{person}{Keping Yang}, {and} \bibinfo{person}{Xiao-ming Wu}.} \bibinfo{year}{2020}\natexlab{}.
\newblock \showarticletitle{M2GRL: A multi-task multi-view graph representation learning framework for web-scale recommender systems}. In \bibinfo{booktitle}{\emph{Proceedings of the 26th ACM SIGKDD international conference on knowledge discovery \& data mining}}. \bibinfo{pages}{2349--2358}.
\newblock


\bibitem[Wang et~al\mbox{.}(2021)]%
        {LWCKD}
\bibfield{author}{\bibinfo{person}{Yuening Wang}, \bibinfo{person}{Yingxue Zhang}, {and} \bibinfo{person}{Mark Coates}.} \bibinfo{year}{2021}\natexlab{}.
\newblock \showarticletitle{Graph structure aware contrastive knowledge distillation for incremental learning in recommender systems}. In \bibinfo{booktitle}{\emph{CIKM}}. \bibinfo{pages}{3518--3522}.
\newblock


\bibitem[Wang et~al\mbox{.}(2023)]%
        {PIW}
\bibfield{author}{\bibinfo{person}{Yuening Wang}, \bibinfo{person}{Yingxue Zhang}, \bibinfo{person}{Antonios Valkanas}, \bibinfo{person}{Ruiming Tang}, \bibinfo{person}{Chen Ma}, \bibinfo{person}{Jianye Hao}, {and} \bibinfo{person}{Mark Coates}.} \bibinfo{year}{2023}\natexlab{}.
\newblock \showarticletitle{Structure aware incremental learning with personalized imitation weights for recommender systems}. In \bibinfo{booktitle}{\emph{AAAI}}. \bibinfo{pages}{4711--4719}.
\newblock


\bibitem[Wu et~al\mbox{.}(2022)]%
        {inmo}
\bibfield{author}{\bibinfo{person}{Yunfan Wu}, \bibinfo{person}{Qi Cao}, \bibinfo{person}{Huawei Shen}, \bibinfo{person}{Shuchang Tao}, {and} \bibinfo{person}{Xueqi Cheng}.} \bibinfo{year}{2022}\natexlab{}.
\newblock \showarticletitle{Inmo: A model-agnostic and scalable module for inductive collaborative filtering}. In \bibinfo{booktitle}{\emph{SIGIR}}. \bibinfo{pages}{91--101}.
\newblock


\bibitem[Xia et~al\mbox{.}(2008)]%
        {xia2008list-wise}
\bibfield{author}{\bibinfo{person}{Fen Xia}, \bibinfo{person}{Tie-Yan Liu}, \bibinfo{person}{Jue Wang}, \bibinfo{person}{Wensheng Zhang}, {and} \bibinfo{person}{Hang Li}.} \bibinfo{year}{2008}\natexlab{}.
\newblock \showarticletitle{Listwise approach to learning to rank: theory and algorithm}. In \bibinfo{booktitle}{\emph{ICML}}.
\newblock


\bibitem[Xia et~al\mbox{.}(2022)]%
        {xia2022device}
\bibfield{author}{\bibinfo{person}{Xin Xia}, \bibinfo{person}{Hongzhi Yin}, \bibinfo{person}{Junliang Yu}, \bibinfo{person}{Qinyong Wang}, \bibinfo{person}{Guandong Xu}, {and} \bibinfo{person}{Quoc Viet~Hung Nguyen}.} \bibinfo{year}{2022}\natexlab{}.
\newblock \showarticletitle{On-Device Next-Item Recommendation with Self-Supervised Knowledge Distillation}. In \bibinfo{booktitle}{\emph{SIGIR}}.
\newblock


\bibitem[Xu et~al\mbox{.}(2020)]%
        {GraphSAIL}
\bibfield{author}{\bibinfo{person}{Yishi Xu}, \bibinfo{person}{Yingxue Zhang}, \bibinfo{person}{Wei Guo}, \bibinfo{person}{Huifeng Guo}, \bibinfo{person}{Ruiming Tang}, {and} \bibinfo{person}{Mark Coates}.} \bibinfo{year}{2020}\natexlab{}.
\newblock \showarticletitle{Graphsail: Graph structure aware incremental learning for recommender systems}. In \bibinfo{booktitle}{\emph{CIKM}}. \bibinfo{pages}{2861--2868}.
\newblock


\bibitem[Yang et~al\mbox{.}(2019)]%
        {yang2019adaptive}
\bibfield{author}{\bibinfo{person}{Yang Yang}, \bibinfo{person}{Da-Wei Zhou}, \bibinfo{person}{De-Chuan Zhan}, \bibinfo{person}{Hui Xiong}, {and} \bibinfo{person}{Yuan Jiang}.} \bibinfo{year}{2019}\natexlab{}.
\newblock \showarticletitle{Adaptive deep models for incremental learning: Considering capacity scalability and sustainability}. In \bibinfo{booktitle}{\emph{KDD}}. \bibinfo{pages}{74--82}.
\newblock


\bibitem[Ying et~al\mbox{.}(2018)]%
        {ying2018graph}
\bibfield{author}{\bibinfo{person}{Rex Ying}, \bibinfo{person}{Ruining He}, \bibinfo{person}{Kaifeng Chen}, \bibinfo{person}{Pong Eksombatchai}, \bibinfo{person}{William~L Hamilton}, {and} \bibinfo{person}{Jure Leskovec}.} \bibinfo{year}{2018}\natexlab{}.
\newblock \showarticletitle{Graph convolutional neural networks for web-scale recommender systems}. In \bibinfo{booktitle}{\emph{Proceedings of the 24th ACM SIGKDD international conference on knowledge discovery \& data mining}}. \bibinfo{pages}{974--983}.
\newblock


\bibitem[Zeng et~al\mbox{.}(2022)]%
        {CL-DRD}
\bibfield{author}{\bibinfo{person}{Hansi Zeng}, \bibinfo{person}{Hamed Zamani}, {and} \bibinfo{person}{Vishwa Vinay}.} \bibinfo{year}{2022}\natexlab{}.
\newblock \showarticletitle{Curriculum Learning for Dense Retrieval Distillation}. In \bibinfo{booktitle}{\emph{SIGIR}}. \bibinfo{pages}{1979–1983}.
\newblock


\bibitem[Zhu et~al\mbox{.}(2023)]%
        {zhu2023reloop2}
\bibfield{author}{\bibinfo{person}{Jieming Zhu}, \bibinfo{person}{Guohao Cai}, \bibinfo{person}{Junjie Huang}, \bibinfo{person}{Zhenhua Dong}, \bibinfo{person}{Ruiming Tang}, {and} \bibinfo{person}{Weinan Zhang}.} \bibinfo{year}{2023}\natexlab{}.
\newblock \showarticletitle{ReLoop2: Building Self-Adaptive Recommendation Models via Responsive Error Compensation Loop}. In \bibinfo{booktitle}{\emph{KDD}}.
\newblock


\bibitem[Zhu et~al\mbox{.}(2020)]%
        {zhu2020ensembled}
\bibfield{author}{\bibinfo{person}{Jieming Zhu}, \bibinfo{person}{Jinyang Liu}, \bibinfo{person}{Weiqi Li}, \bibinfo{person}{Jincai Lai}, \bibinfo{person}{Xiuqiang He}, \bibinfo{person}{Liang Chen}, {and} \bibinfo{person}{Zibin Zheng}.} \bibinfo{year}{2020}\natexlab{}.
\newblock \showarticletitle{Ensembled CTR Prediction via Knowledge Distillation}. In \bibinfo{booktitle}{\emph{CIKM}}.
\newblock


\bibitem[Zhu et~al\mbox{.}(2024)]%
        {Zhu_2024}
\bibfield{author}{\bibinfo{person}{Yaochen Zhu}, \bibinfo{person}{Liang Wu}, \bibinfo{person}{Qi Guo}, \bibinfo{person}{Liangjie Hong}, {and} \bibinfo{person}{Jundong Li}.} \bibinfo{year}{2024}\natexlab{}.
\newblock \showarticletitle{Collaborative Large Language Model for Recommender Systems}. In \bibinfo{booktitle}{\emph{Proceedings of the ACM on Web Conference 2024}} \emph{(\bibinfo{series}{WWW ’24})}. \bibinfo{publisher}{ACM}.
\newblock
\urldef\tempurl%
\url{https://doi.org/10.1145/3589334.3645347}
\showDOI{\tempurl}


\end{thebibliography}

\pagebreak
\newpage
\clearpage
\appendix
\nobalance
\section{Appendix}
The source code of CCD is publicly available through the author’s GitHub repository.\footnote{\url{https://github.com/Gyu-Seok0/CCD_KDD24}}
\subsection{List-wise distillation}
\label{APP:KD}
The list-wise distillation \cite{DERRD, hetcomp, DCD} trains the student to emulate the item permutation (i.e., ranking orders) predicted by the teacher.
Specifically, they define a probability for observing each permutation based on the Plackett-Luce model \cite{marden1996analyzing}, then train the student to maximize the likelihood of the teacher permutations \cite{xia2008list-wise}.
For each user $u$, the item ranking list predicted by \tRS is denoted by $\pi^T_u$.
The list-wise KD loss is defined as the negative log-likelihood of permutation probability \cite{hetcomp}:
\begin{align}
    \mathcal{L}_{KD}= -\sum_{u \in \mathcal{U}} \log p(\pi^T_u\mid S) = -\sum_{u \in \mathcal{U}} \log \prod_{n=1}^{N} \frac{\exp \left(\hat{r}_{u,\pi^T_u(n)}^S\right)}{\sum_{i = n}^{|\mathcal{I}|} \exp \left(\hat{r}_{u,\pi^T_u(i)}^S\right)}.
\end{align}
$\pi^T_u(n)$ denotes the $n$-th item in $\pi^T_u$. $N$ is a hyperparameter reflecting the length of the recommendation list.
In this work, we set $N=50$ \cite{hetcomp}.
By minimizing the loss, the student model learns to preserve the detailed ranking orders of top-$N$ items in $\pi^T_u$, while penalizing the remaining items below the $N$-th rank.

This distillation approach has shown remarkable performance in many applications such as recommendation \cite{DERRD, DCD, hetcomp, Kang_unbiased} and document retrieval \cite{reddi2021rankdistil, CL-DRD}.
Although more complicated variants have been studied, we obtained satisfactory results with the default distillation loss.
We assessed the effectiveness of this distillation in our experimental setup, and the results are presented in Table \ref{tab:KD_result}.

\subsection{Experiment details}
\label{App:setup}
\subsubsection{\textbf{Dataset.}} 
We use two public datasets with real-world user-item interactions: Gowalla\footnote{https://snap.stanford.edu/data/loc-gowalla.html} and Yelp\footnote{https://www.yelp.com/dataset}. We filter out users and items with fewer than 10 interactions in both datasets. 
The total data statistics after preprocessing are summarized in Table \ref{tab:dataset}.
\begin{table}[h]
\caption{Dataset statistics.}
\centering
\resizebox{0.9\columnwidth}{!}{
\begin{tabular}{ccccc}
\hline
\textbf{Dataset}          & \textbf{\#Users}               & \textbf{\#Items} & \textbf{\#Interactions} & \textbf{Sparsity(\%)} \\ \hline\hline
\textbf{Gowalla} & 29,858                         & 40,988           & 1,027,464               & 99.91                  \\ 
\textbf{Yelp}    & \cellcolor[HTML]{FFFFFF}14,950 & 12,261           & 342,617                 & 99.81                       \\\hline
\end{tabular}
}
\label{tab:dataset}
\end{table}


\subsubsection{\textbf{Teacher and student configuration.}} 
\label{App:teacher_setup}
We provide the detailed configuration of the teacher and the student used in the experiments.
Table \ref{tab:KD_result} presents the number of learning parameters as well as the effectiveness of distillation in our setup.

\smallsection{Teacher.}
Following the teacher configuration of the recent KD work \cite{hetcomp}, we construct a massive teacher system by ensembling multiple large models.
Initially, we evaluate the effectiveness of five different recommendation models for each dataset: matrix factorization (MF) \cite{BPR}, metric learning \cite{CML}, deep neural network \cite{NeuMF}, graph neural network (GNN) \cite{he2020lightgcn}, and variational autoencoder (VAE) \cite{VAE}. 
Subsequently, we select up to two models for each dataset that demonstrate high performance. 
Lastly, we incrementally increase the capacity of the teacher system by augmenting the number of models or dimension sizes until its performance no longer improves. 
For Gowalla, we construct the teacher system using five MF-based models \cite{BPR} and two VAE-based models \cite{VAE}, each initialized with distinct random seeds and set to 64 dimensions. 
For Yelp, we employ five GNN-based models \cite{he2020lightgcn}, each initialized with distinct random seeds and set to 128 dimensions. 
We adopt the ensemble scheme utilized in \cite{hetcomp}.

\smallsection{Student.}
For the student, we employ two backbone models: MF-based model \cite{BPR} and GNN-based model \cite{he2020lightgcn}.
We set a small embedding size for the student (16 for Gowalla and 8 for Yelp), considering the teacher size for each dataset.

\smallsection{Effectiveness of KD.}
We assessed the effectiveness of the list-wise KD technique (\ref{APP:KD}) on our teacher-student setup.
The recommendation performance after the distillation is presented in Table~\ref{tab:KD_result}.

\begin{table}[ht!]
\centering
\caption{Recall@20 results after distillation for each data block. \#Params. denotes the number of parameters.}
\label{tab:KD_result}
\resizebox{0.95\linewidth}{!}{
\begin{tabular}{c|c|cccc|c}\hline
\textbf{Dataset} &  & $D_1$ & $D_2$ & $D_3$ & $D_4$ & \multicolumn{1}{c}{\textbf{\#Params.}} \\ \hline \hline
 & \textbf{Teacher} & 0.1849 & 0.1708 & 0.1698 & 0.1768 & 33.47M \\
\multirow{-2}{*}{\textbf{Gowalla}} & \textbf{Student} & 0.1822 & 0.1686 & 0.1652 & 0.1677 & 1.13M \\ \hline
 & \textbf{Teacher} & 0.1237 & 0.1117 & 0.1244 & 0.1195 & 18.24M \\
\multirow{-2}{*}{\textbf{Yelp}} & \textbf{Student} & 0.1161 & 0.1138 & 0.1256 & 0.1147 & 0.22M \\ \hline
\end{tabular}}
\end{table}




\end{document}